\documentstyle[10pt,aaspp4]{article}
\def\simg{\mathrel{%
      \rlap{\raise 0.511ex \hbox{$>$}}{\lower 0.511ex \hbox{$\sim$}}}}
\def\siml{\mathrel{%
      \rlap{\raise 0.511ex \hbox{$<$}}{\lower 0.511ex \hbox{$\sim$}}}}
\date{today}

\begin{document}
\begin{center}
\title{SIMULATIONS OF GAMMA-RAY BURSTS FROM EXTERNAL SHOCKS: \\
TIME VARIABILITY AND SPECTRAL CORRELATIONS}
\author{A. Panaitescu and P. M\'esz\'aros\altaffilmark{1}} 
\affil{Department of Astronomy \& Astrophysics, \\
Pennsylvania State University, University Park, PA 16802}
\altaffiltext{1}{also Center for Gravitational Physics and Geometry,
Pennsylvania State University}
\bigskip
\end{center}

\begin{abstract}

We compute burst spectra and time structures arising from synchrotron and 
inverse Compton scattering by non-thermal electrons accelerated in shocks 
which form during the interaction between a thin ultra-relativistic fireball 
and a stationary external medium. We investigate the effect of varying the 
most important model parameters on the resulting burst spectra, and we
present a set of correlations among the spectral and temporal features of
the bursts. The spectral hardness, various spectral-temporal correlations and
the spectral evolution of the simulated bursts are compared to those of 
observed bursts for a representative set of model parameters. Multi-pulse
structures are simulated using a variable magnetic field and anisotropic
emission, and the most important spectral and temporal properties of these 
pulses are compared with observations.

\end{abstract}


\section{Introduction}

The observed isotropy (\cite {mee92}, \cite{briggs96}), inhomogeneity (as 
shown by $\langle V/V_{max} \rangle$) and the deviation from a $-1.5$ slope 
power-law of the $\log N-\log P$ distribution for the fainter bursts 
(\cite{mee92}, \cite{hor94}) provide string support for the hypothesis that
Gamma-Ray Bursts (GRBs) are of cosmological origin. Other observations, 
such as spectral hardness--brightness correlations (\cite{mit92a}, 
\cite{pac92}), spectral hardness--duration anti-correlation (\cite{kou93}), 
and the possible time dilation and duration--brightness anti-correlation 
(\cite{nor94}, \cite{nor95}; see however \cite{mit96}), while more equivocal, are also 
generally compatible with this hypothesis.  The large energy that the 
cosmological source must release  suggests that relativistic effects are 
likely involved in GRBs. In this paper we consider bursts that arise when an
ultra-relativistic cold shell (fireball) is decelerated by interaction with
the interstellar medium, or a pre-ejected slow wind. As a result of the 
deceleration an ultra-relativistic blast wave (``forward shock" -- FS) 
propagates into the external medium (EM), transferring a substantial part 
of the fireball kinetic energy to the shocked EM, while another shock (``reverse 
shock" -- RS) propagates back into the fireball. This is
the generic model usually referred to as the ``external shock model" 
(\cite{mesz93a}). 
If the shell is in the linear broadening regime before it is substantially decelerated by the EM
(as described by M\'esz\'aros, Laguna \& Rees 1993), 
a situation that is expected under a wide range of conditions,
then the RS is quasi-newtonian and therefore less efficient than the blast wave in 
converting the shell's kinetic energy into heat.
In another very likely scenario (the ``internal shock model" -- \cite{rees94}, \cite{paxu94}) 
the energy conversion 
takes place when several ultra-relativistic shells collide with each other, 
before the deceleration caused by the EM becomes important. 
Here we focus on the first model.

In order to simulate the propagation of the two shocks and to model the 
fireball--EM interaction at large Lorentz factors ($\Gamma_0 \geq 100$),
we have developed a one-dimensional hybrid (finite differencing 
+ exact Riemann solver) hydrodynamic code (\cite{wen96}). As the 
conversion of kinetic to internal energy takes place, the heat stored in the 
post-shock gas can be released as radiation, generating a burst. In a previous 
paper (\cite{alin97}) we have simulated burst light-curves from fireballs 
with moderate Lorentz factors ($\Gamma_0 \leq 200$), using a simplified prescription 
for the energy release. The results were single-hump bolometric light-curves 
with a large temporal asymmetry (light-curve decay lasting substantially longer 
than its rise), practically insensitive to variations of the EM density. 
In order to carry out an appropriate comparison of this model
with the rich observational database that has been accumulated by BATSE and
other experiments, we need to compute the spectra of such bursts, and to study 
the burst spectral evolution and its correlation with the other observational 
properties and parameters of the model, by taking into
consideration specific energy release mechanisms. The spectral hardness of the 
observed bursts and its time evolution are well studied, and to reproduce these
should represent a major goal of any GRB model. In this work, we calculate the
effect of the source evolution on the burst spectrum, and explore the
spectral-temporal correlations predicted by the model. We also explore
the physical requirements necessary in order for this model to produce 
multiple-humped light curves, and discuss the possible limitations. 
Spherical symmetry is assumed for simplicity throughout the paper, which
also describes well the case of jets with an opening angle $\theta > 
\Gamma_0^{-1}$. The importance of a non-planar symmetry can be assessed 
from the shape of the light-curves and pulses presented below.

\section{Model Parameters, Assumptions, Approximations, Scaling Relations}

 The most important parameters that describe the dynamics of the fireball-EM
interaction and the energy release mechanism are listed in Table 1, together
with most relevant equations in which they appear.
The evolution of an impulsive fireball has two phases: a free expansion phase, 
when the amount of swept up EM is small and the deceleration caused by 
it can be neglected; and a decelerated expansion phase, when the fireball 
kinetic energy is used to heat the swept-up EM. The fireball dynamics during 
the first stage was calculated analytically and simulated numerically by 
\cite{mesz93b}. The evolution during this stage is determined by three 
parameters: (1) the energy $E_0 = 10^{51}\,E_{0,51}\; {\rm ergs}$ deposited in the 
ejected fireball; (2) the entrained baryonic mass $M$, parameterized through the
dimensionless entropy $\Gamma_0=E_0/Mc^2 \gg 1$; and (3) the initial size of the 
fireball $r_0$ (which may be of the order of the neutron star radius , $r_0 
\simg 10^6 {\rm cm}$). At the beginning of the free expansion phase the 
fireball is accelerated as the radiation energy $E_0$ contained in it is 
adiabatically transformed into bulk motion energy, and becomes stretched out 
into a thin shell. The absence of a strong burst precursor with a 
quasi-thermal spectrum suggests that most of the initial internal fireball 
energy is in the form of baryonic kinetic energy when the shell becomes 
optically thin and photons escape from it. Therefore, the maximum Lorentz factor
attained by the fireball is $\siml \Gamma_0$, corresponding to a kinetic energy
$\siml E_0$. 

\begin{table}[ht]
\begin{center}
TABLE 1 \\
{\sc Summary of the most important parameters and physical quantities that characterize
the dynamics of the interaction fireball--EM and the burst energy release} \\ [4 ex]
\begin{tabular}{ccc} \hline \hline
\rule[-4mm]{0mm}{10mm} Symbol & Definition & Equation \\ \hline
\rule[-2mm]{0mm}{6mm} $E_0$ & initial fireball kinetic energy & --  \\
\rule[-2mm]{0mm}{6mm} $\Gamma_0$ & initial fireball Lorentz factor & $E_0=\Gamma_0 M c^2$ \\
\rule[-2mm]{0mm}{6mm} $t_{dec}$ & hydrodynamic time-scale & (\ref{tdec}) \\ \hline
\rule[-2mm]{0mm}{6mm} $\lambda_B$ & magnetic field parameter & $U_B = \lambda_B U_{int}$ \\
\rule[-2mm]{0mm}{6mm} $B$ & magnetic field intensity & (\ref{Bmag}) \\
\rule[-2mm]{0mm}{6mm} $\kappa$ & electron-to-proton energy ratio & (\ref{kappa}) \\ \hline
\rule[-2mm]{0mm}{6mm} $\gamma_m$ & minimum electron Lorentz factor & (\ref{gammamin}) \\
\rule[-2mm]{0mm}{6mm} $\varepsilon_{SY/IC}$ & co-moving SY/IC photon energy & (\ref{sy}),(\ref{ic}) \\
\rule[-2mm]{0mm}{6mm} $E^{SY/IC}$ & detector SY/IC photon energy &
                       (\ref{SYRS}),(\ref{ICRS}),(\ref{SYFS}),(\ref{ICFS}) \\
\rule[-2mm]{0mm}{6mm} $Y^{RS/FS}$ & Kompaneets parameter & (\ref{CompRS}),(\ref{CompFS}),(\ref{CompKN}) \\
\rule[-2mm]{0mm}{6mm} $t^{SY}$ & synchrotron cooling time-scale & (\ref{cool}) \\ \hline
\end{tabular}
\end{center}
\end{table}

After the fireball Lorentz factor attains its 
maximum value the fireball coasts at constant $\Gamma_0$ and later reaches 
optical thinness. The deceleration caused by the interaction with the EM must 
be taken into account when the energy stored into the shocked EM is a 
substantial fraction of the initial kinetic energy $E_0$. The shocked EM internal 
energy is much larger than its rest mass energy, since its random (or thermal) 
Lorentz factor is $\sim \Gamma_0 \gg 1$. Throughout most of this paper we assume that the 
EM is homogeneous, characterized by a single parameter: its number density 
$n = 1\, n_0\; {\rm cm}^{-3}$.
The deceleration time-scale in the stationary frame (with respect to the Earth)
of the center of explosion (the laboratory frame) is
\begin{equation}
t_{dec} = r_{dec}/c \simeq (E_0/\Gamma_0^2 n m_p c^5)^{1/3}
\simeq 8.3\times 10^5 \; E_{0,51}^{1/3}\, n_0^{-1/3}\, \Gamma_{0,2}^{-2/3}~{\rm s}~,
\label{tdec}
\end{equation}
where $\Gamma_0=10^2\,\Gamma_{0,2}$. Due to the relativistic motion of the 
source, the stationary observer receives radiation emitted in $dt$ in a much 
shorter time $dT=dt/[2\,\Gamma^2(t)]$ (\cite{rees66}), where $\Gamma(t) < \Gamma_0$ is 
the Lorentz factor of the shocked emitting medium. The burst duration is then
approximately
\begin{equation}
T_b \approx 10\,t_{dec}\,/\,2\,\Gamma_0^2 = 420 \;\, E_{0,51}^{1/3}\,
n_0^{-1/3}\, \Gamma_{0,2}^{-8/3}\;  {\rm s}\; ,
\label{duration}
\end{equation}
where a factor of 10 was included in order to account for the progressive 
decrease of the flow Lorentz factor of the radiating medium. Equation 
(\ref{duration}) and the observed GRB durations imply that $100 \siml \Gamma_0 
\siml {\rm few} \times 1000$. It also shows that the burst peak flux $F_p$
satisfies $F_p (\sim E_0 D^{-2} T_b^{-1}) \propto E_0^{2/3}\,\Gamma_0^{8/3}\, 
n^{1/3}\, D^{-2}$, where $D$ is 
the distance to source, if most of the available energy $E_0$ is radiated.

The dynamics and energetics  of the deceleration phase were calculated by 
\cite{rees92} and by \cite{sari95}. For computational efficiency, the numerical 
simulations presented here were started from $0.5\,t_{dec}$, when only 
$\simeq 12\%$ of the EM mass within $1\,r_{dec}$ had been swept up and the 
deceleration prior to this time can be safely neglected. At $t=0.5\,t_{dec}$ the only physical
parameter that depends on $r_0$ is the internal pressure $P$ of the fireball, 
and in fact this pressure is irrelevant as long as the shell is cold ($P \ll 
\rho c^2$, $\rho$ is rest mass density). Therefore, the hydrodynamics of the 
shell--EM collision is characterized by the set of three parameters 
($E_0,\Gamma_0,n$). 

In the co-moving frame, the shocked EM has typical densities $\sim 10^3\,{\rm particles/cm^3}$, and 
can radiate away its internal energy through synchrotron radiation (SY) and 
inverse Compton (IC) scattering of the SY photons, in the presence of a 
modest magnetic field. Such mechanisms were considered by \cite{mesz94},
who studied the spectral properties of bursts arising from external shocks, 
and by Sari, Narayan \& Piran (1996), who derived constraints on the radiation 
mechanisms parameters from the variability observed in most bursts and from 
efficiency considerations. The galactic magnetic field, even when compressed 
behind the FS, would be too weak to lead to efficient radiation. However, a
frozen-in magnetic field present in the fireball (and thus in the fluid behind the RS)
would allow the post-FS 
material to cool by IC scattering of the SY photons coming from the post-RS 
medium. The swept-up EM could radiate even more efficient if a random turbulent 
magnetic field builds up in it. In our calculation we use for simplicity this
latter scenario; moreover, a frozen-in magnetic field will usually have only a 
fraction of the strength of a turbulent magnetic field at equipartition 
(when the magnetic field energy density is equal to the internal energy density of the gas) .

We use the following assumptions and approximations in order to simulate the 
emission of SY and IC photons from the gas behind the two shocks: 

[1] the magnetic field $B$ is parameterized relative to the internal 
energy density $U_{int}$: $U_B=\lambda_B\, U_{int}$, where $U_B=B^2/\,8\pi$ is the magnetic field
energy density. For strong shocks, equation (8) derived by 
\cite{bla76} yields $U_{int} = 3 \times 10^{-3}\, \,n_0\,\Gamma_{FS}^2\; 
{\rm erg/cm}^3$, where $\Gamma_{FS}$ is the Lorentz factor of the FS shock, so
that $B = 0.27\,\lambda_B^{1/2}\, n_0^{1/2}\,\Gamma_{FS}\;{\rm G}$. Since the 
post-shock fluids are very close to hydrostatic equilibrium, $U_{int}$, and 
therefore $B$, have almost the same values behind both shocks.

[2] shock acceleration leads to a power-law distribution of electrons 
\begin{equation}
d{\cal N}_e(\gamma_e)= C \gamma_e^{-p}\,d\gamma_e \;, \quad \gamma_m \leq 
\gamma_e \leq \gamma_M \,, 
\end{equation}
where $\gamma_e$ is the random electron Lorentz factor and
${\cal N}_e$ is the number density of electrons.
Such a distribution is initialized in every grid cell after it is swept up by 
one of the two shocks, and its subsequent evolution is determined solely by 
the SY and IC losses. Therefore we do not take into account adiabatic losses or
further energy exchange between protons and electrons. The former simplification
is justified by the fact that the electron cooling time-scale is much lower
than the dynamic time-scale (as shown below). We have taken 
$\gamma_M/\gamma_m=10$ because the cooling time-scales for larger $\gamma_M$ 
would be too short ($\siml 10^{-5}\,t_{dec}$, typically) and it would require a large 
computational effort to follow accurately the evolution of these very 
energetic electrons. The effect of a larger ratio $\gamma_M/\gamma_m$ on the burst 
spectrum can be easily estimated in the figures presented in the next section.
Moreover, if $\Gamma_0$ is not low ($\approx 100$) or if the magnetic 
field is not weak or the shock acceleration inefficient (i.e$.$ low $\gamma_m$), 
the most energetic electrons radiate at energies above the upper limit of the 
BATSE window (10 keV -- few MeV). The power-law index $p$ was chosen to be 3. 

[3] The minimum electron random Lorentz factor$\gamma_m$ is determined by a 
parameter $\kappa$ which is the ratio of 
the energy in electrons and that in the monoenergetic protons (with Lorentz 
factor $\gamma_p$), after shock acceleration: 
\begin{equation}
\int_{\gamma_m}^{\gamma_M} 
d{\cal N}_e(\gamma_e)\,m_e\,(\gamma_e-1)=\kappa\,n_p\,m_p\,(\gamma_p-1) \, ,
\label{kappa}
\end{equation} 
where $n_p$ is the density of protons. Using the equality of the sum of the 
electronic and protonic partial pressures and the total pressure (determined by 
the hydrodynamics of the fireball--EM interaction), $\gamma_m$ and the electron 
distribution are completely determined: 
\begin{equation}
\gamma_m (\kappa) = 3\,[(p-2)/(p-1)]\,
[(1-X^{1-p})/(1-X^{2-p})]\,f_{\kappa}\,(m_p/m_e)\,(P/\rho c^2) \,,
\label{gammamin}
\end{equation} 
where $X=\gamma_M/\gamma_m$ and $f_{\kappa}=\kappa/(\kappa+1)$. 
This result is valid for $\gamma_p \gg 1$ (equivalent with $P \gg \rho c^2$,
which is true for the fluid behind the FS); a similar result can be obtained in the limit
$\gamma_p -1 \ll 1$ (i.e$.$ $P \ll \rho c^2$, which is correct for the fluid behind the RS). 
For the ultra-relativistic FS, 
equations (8)--(10) from \cite{bla76} lead to $P/\rho c^2 = 0.24 \,\Gamma_{FS}$ 
and $\gamma_{m,FS} = 660\,f_{\kappa}\,\Gamma_{FS}$. \cite{mesz93b} have 
shown that the evolution of the fireball thickness $\Delta$ during the free 
expansion phase and for $r > r_0\, \Gamma_0^2 = 10^{10}\, r_{0,6}\, \Gamma_{0,2}^2 
\; {\rm cm}$ is $\Delta = r\,/\,\Gamma_0^2$. This leads to a fireball density 
at $r \sim r_{dec}$ that is much larger than that of the EM and produces a mildly relativistic RS. 
Numerically, we found that the Lorentz factor of the RS in the frame of the yet
un-shocked fluid is practically independent of $\Gamma_0$: $\Gamma_{RS} \simeq 
1.1$ . In this case it can be shown that $P/\rho c^2 \simeq 4 \times 10^{-2}$
which leads to $\gamma_{m,RS} \simeq 100\,f_{\kappa}$.

[4] In the co-moving frame, the SY radiation emitted by any electron is approximated as 
monochromatic, with a frequency equal to the peak frequency $\nu_c$ (averaged over the
pitch angle) of the SY spectrum emitted by an electron with Lorentz
factor $\gamma_e(t)$ (that evolves in time, as the electron loses energy): 
\begin{equation}
\varepsilon_{SY} \simeq 4.0\times 10^{-9}\,\gamma_e^2(t)\,B\;{\rm eV} \;.
\label{sy}
\end{equation}
The electron cooling and continuous electron injection will produce spectra 
that are flatter than the spectrum of the SY radiation emitted by a single
electron below and above the peak frequency $\nu_c$ ($\nu F_{\nu} \propto \nu^{4/3}$
and $\nu F_{\nu} \propto \nu^{3/2} \exp{[-\nu/\nu_c]}$, respectively), so that
the effect of integrating over time and over electron distribution hinders
the features of a single electron spectrum. Thus, this approximation is in fact
better than it seems at first sight.

[5] The spectrum of the SY photons up-scattered in the Thomson regime is also approximated as 
monochromatic, at the average energy of the IC spectrum for $\gamma_e(t)$: 
\begin{equation}
\varepsilon_{IC} = 4/3\,\gamma_e^2(t)\,\varepsilon_{SY} \;.
\label{ic}
\end{equation} 
The Klein-Nishina (K-N) effect on the scattering of SY photons with energies comparable or larger 
than $m_e c^2/\gamma_e(t)$ is taken into account. The SY energy density 
$U_{SY}$, necessary for calculating the IC losses, is computed as an integral 
over the volume of the shocked media of the SY local output. There is a strong 
relativistic beaming of the SY photons due to the radial motions of the origin
of a given photon and the place where the scattering takes place: as seen from the 
co-moving frame of the up-scattering region, the SY source is moving away, 
unless the two regions (of SY emission and of IC scattering) are moving 
in the same radial direction.
We assumed that the $U_{SY}$ spectrum is monochromatic, at the peak frequency of 
the SY spectrum generated by the most numerous (and least energetic) electrons 
that are in the same volume element where the IC scattering takes place. This 
approximation is justified to some extent by the aforementioned strong 
relativistic beaming and the geometrical dilution of the SY output, which 
should make the contribution to the $U_{SY}$ of the SY emission from the 
vicinity of IC scattering place to be dominant. Due to this assumption the 
IC spectra shown in the next section are calculated using only the following 
combinations: (i) SY-RS photons scattered on electrons accelerated by the RS 
and (ii) SY-FS photons scattered on electrons accelerated by the FS. The mixed 
combinations (iii) SY-RS photons scattered by FS electrons and (iv) SY-FS photons 
scattered by RS electrons are not taken into account. We will assess the effect on the 
computed spectra of neglecting the last two combinations.

The approximation of taking the SY and IC spectra of a single electron and that of the SY photon field
to be up-scattered as
monochromatic is done for computational efficiency.  The results presented in
the next section were obtained from numerical runs that last few hours on a
Sparc Sun 20 Station for the lowest $\Gamma_0$ considered there ($\Gamma_0=100$)
and up to few days for the highest $\Gamma_0$ we used ($\Gamma_0=800$). Most of
this computational effort is used to calculate the burst spectrum  
by integrating over the volume of the shocked
fluid (which reduces to integrating over the radial coordinate and the angle
relative to the line of sight toward the center of symmetry) and over the electron
distribution in each infinitesimal volume element, and repeating this triple
integral after a time short enough to accurately treat the evolution of
the most energetic electrons (which have the shortest cooling time-scale). Adding
another integral within the triple integral, in order to include the real SY or
IC spectrum from a single electron, would lead to excessively long runs.

The above analytic considerations and approximations allow us to calculate the energy $E_p$ at 
the peak of the power per logarithmic energy interval ($\nu F_{\nu}$) for the SY and IC 
spectra from both shocks, as seen from the detector frame. Numerically, we found that
about 50\% of the total energy released by a burst is emitted from $t=1\,t_{dec}$
until $t=1.5\,t_{dec}$. During this time $\Gamma_{FS}$ decreases from $\simeq 0.6\,\Gamma_0$
to $\simeq 0.4\,\Gamma_0$, therefore, to a good approximation,
$\Gamma_{FS} \simeq \Gamma_0/2$, so that 
\begin{equation}
\gamma_{m,FS} \simeq 1.3 \times 10^5\,f_{\kappa}\,(\Gamma_0/400) 
\label{gammaFS}
\end{equation}
and
\begin{equation}
B \simeq 54\;\lambda_B^{1/2}\,n_0^{1/2}\,(\Gamma_0/400)\; {\rm G} \;.  
\label{Bmag}
\end{equation}
Taking into account that the relativistic motion of the radiating fluid boosts
the co-moving energy by a factor between $\Gamma$ (if the fluid moves at an angle
$\Gamma^{-1}$ from the line of sight toward the fireball's center) and $2\,\Gamma$
(if the fluid moves on this line of sight), where $\Gamma \simeq 0.7\,\Gamma_{FS}$
is the flow Lorentz factor of the shocked fluid, we obtain for the SY-RS radiation:
\begin{equation}
E_p^{SY,RS} \simeq 0.4\, f_{\kappa}^2\, \lambda_B^{1/2}\, n_0^{1/2}\, (\Gamma_0/400)^2\;{\rm eV} \;. 
\label{SYRS}
\end{equation}
The co-moving energy of the SY-RS photons $\varepsilon_{SY,RS}=2 \times 10^{-3}\,
f_{\kappa}^2\,\lambda_B^{1/2}\,n_0^{1/2}\,(\Gamma_0/400)\; {\rm eV}$ is well below the limit for K-N
scattering $m_e c^2/\gamma_{m,RS} = 5\,f_{\kappa}^{-1}\; {\rm keV}$, therefore 
\begin{equation}
E_p^{IC,RS} \simeq 6\, f_{\kappa}^4\, \lambda_B^{1/2}\, n_0^{1/2}\, (\Gamma_0/400)^2\;{\rm keV}\; .
\label{ICRS}
\end{equation}
The SY photons emitted by post-FS electrons with Lorentz factor $\gamma_{m,FS}$ arrive at detector at
\begin{equation}
E_p^{SY,FS} \simeq 800\, f_{\kappa}^2\, \lambda_B^{1/2}\, n_0^{1/2}\, (\Gamma_0/400)^4\;{\rm keV} \;,
\label{SYFS}
\end{equation}
and, in the co-moving frame, are too energetic to be up-scattered in the Thomson regime:
$\varepsilon_{SY,FS} = 4\, f_{\kappa}^2\, \lambda_B^{1/2}
\, n_0^{1/2}\, (\Gamma_0/400)^3\; {\rm keV} \gg m_e\,c^2/\gamma_m = 4\,
f_{\kappa}^{-1}\, (\Gamma_0/400)^{-1} \; {\rm eV}$, as long as 
\begin{equation}
\label{cond}
\log\Gamma_{0,2}+ \frac{3}{4}\log f_{\kappa}+\frac{1}{8}\log\lambda_B +
\frac{1}{8}\log n_0 \simg -0.15 \; .
\end{equation}
Let's assume that inequality (\ref{cond}) is satisfied (we argue below that it must be so for an
efficient burst). The energy $m_e c^2/\gamma_e$ is a good measure of the photon energy
above which the K-N reduction is very effective, in the sense that it drastically reduces
the intensity of the IC component. 
The up-scattered radiation will be emitted when 
the FS electrons have cooled enough so that $\varepsilon_{SY,FS}(\gamma_e) 
\leq m_e c^2/\gamma_e$, implying $\gamma_e \leq \gamma_{KN} = 5 \times 10^4\,B^{-1/3}$, 
where $B$ is lower than estimated above (eq$.$[\ref{Bmag}]), as the shocked material has lost some 
internal energy. To a good approximation, this fraction can be taken 1/2, so that 
$\gamma_{KN} = 1.5 \times 10^4\,\lambda_B^{-1/6}\,n_0^{-1/6}\,(\Gamma_0/400)^{-1/3}$.
The SY radiation emitted by FS electrons cold enough to scatter their own SY photons in a mild
K-N regime have a detector frame energy less than
\begin{equation}
E_{KN}^{SY,FS} \simeq 7\,\lambda_B^{1/6}\,n_0^{1/6}\,(\Gamma_0/400)^{4/3}\;{\rm keV} \;,
\label{KN}
\end{equation}
which gives the peak energy of the up-scattered spectrum from the FS:
\begin{equation} 
E_p^{IC,FS} \approx 0.6\, \lambda_B^{-1/6}\, n_0^{-1/6}\, (\Gamma_0/400)^{2/3}\; {\rm TeV}\; .
\label{ICFS}
\end{equation}

The optical depth for Thompson scattering of the shocked fireball is 
$\tau_{RS} = E_0\,\sigma_{Th}\,/\,4\pi\, m_p c^2\, r_{dec}^2\, \Gamma_0 
\simeq 10^{-6}\; E_{0,51}^{1/3}\, n_0^{2/3}\, (\Gamma_0/400)^{1/3} \ll 1$.
The effect of IC scattering on the SY-RS spectrum and on electron cooling
can be assessed through the Kompaneets parameter $Y^{RS} = \gamma_{m,RS}^2\,\tau_{RS}$ :
\begin{equation}
Y^{RS} \simeq 10^{-2}\, f_{\kappa}^2 E_{0,51}^{1/3} n_0^{2/3} (\Gamma_0/400)^{1/3} \ll 1 \, .
\label{CompRS}
\end{equation}
Calculating a similar Kompaneets parameter for the FS is more difficult because earlier
accelerated electrons can be so cold that they scatter their own SY photons in the Thomson
regime while more recently accelerated electrons are very energetic and scatter their SY photons
in the extreme K-N regime. A simple way of obtaining upper limits for this parameter would be
to assume that all electrons have the same random Lorentz factor and that the up-scattering
takes place at the limit between the Thomson and the K-N regimes. (At given energy $\varepsilon_0$
of the incident photon, the Kompaneets parameter for electrons colder than $\gamma_0=m_e c^2/\varepsilon_0$
increases as $\gamma_e^2$ while for electrons with random Lorentz factors above $\gamma_0$
the same parameter increases as $\ln{[2\gamma_e/\gamma_0]}$.) The SY photons
emitted by electrons with $\gamma_{m,FS}$ (eq$.$ [\ref{gammaFS}]) are up-scattered in this mild
K-N regime by electrons that have $\gamma_e = 140\,f_{\kappa}^{-2}\,\lambda_B^{-1/2}\,n_0^{-1/2}\,
(\Gamma_0/400)^{-3}$; for such scatterings the Kompaneets parameter is:
\begin{equation}
Y^{FS} \approx 10^{-5} f_{\kappa}^{-4}\,\lambda_B^{-1}\, E_{0,51}^{1/3}\,n_0^{-1/3}\,
(\Gamma_0/400)^{-20/3} \ll 1 \,.
\label{CompFS}
\end{equation}
Before reaching $\gamma_0$ calculated above, electrons are cold enough to scatter the SY photons they
produce (see equation for $\gamma_{KN}$ above). Assuming again that all electrons are monoenergetic
and have $\gamma_e=\gamma_{KN}$, the Kompaneets parameter is:
\begin{equation}
Y_{KN}^{FS} \approx 10^{-1} \lambda_B^{-1/3}\,E_{0,51}^{1/3}\,n_0^{1/3}\,(\Gamma_0/400)^{-4/3} \,.
\label{CompKN}
\end{equation}  
For electrons colder than $\gamma_{KN}$ the $Y$ parameter should increase as
$\gamma_e^2$ while for more energetic electrons the same parameter should decrease as $\gamma_e^{-4}$.
Thus equation (\ref{CompKN}) gives an upper limit on the expected intensity of the IC-FS component
relative to that of the SY-FS emission.
In deriving equations (\ref{CompFS}) and (\ref{CompKN}) we approximated the mass of the swept up EM 
by a fraction $1/\Gamma_0$ of the fireball mass. 
We can conclude from equations (\ref{CompRS}) and (\ref{CompFS}) that the IC
emission is not expected to alter substantially the intensity of the SY radiation from the two shocks
or the synchrotron cooling time-scale of electrons. 

The energy release mechanisms considered in this model involve only two 
important parameters ($\kappa,\lambda_B$) which, based on equation (\ref{SYFS}),
must satisfy the double inequality 
\begin{equation}
0.1 \siml \log\Gamma_{0,2}+ \frac{1}{2}\log f_{\kappa}+\frac{1}{8}\log\lambda_B +
\frac{1}{8}\log n_0 \siml 0.7 \; ,
\label{window}
\end{equation}
to ensure that the burst fluence in 
the BATSE window corresponds to a significant fraction of the total energy radiated by the source.
In the laboratory frame, the synchrotron cooling time of the least energetic FS electrons is
$t^{SY}= (3\,m_e c\,/\,4\,\sigma_{Th}\,U_B\,\gamma_m)\,\Gamma \simeq
140 \;f_{\kappa}^{-1}\, \lambda_B^{-1}\,
n_0^{-1}\, \left(\,\Gamma_0/400 \,\right)^{-2}\; {\rm s}$, or
\begin{equation}
t^{SY} \simeq 4 \times 10^{-4}\; f_{\kappa}^{-1}\, \lambda_B^{-1}\, n_0^{-2/3}\,
E_{0,51}^{-1/3}\,(\Gamma_0/400)^{-4/3}\; t_{dec}\; .
\label{cool}
\end{equation}
If the SY cooling time-scale is larger than the hydrodynamic time-scale $t_{dec}$, 
the progressive fluid deceleration and the adiabatic cooling of the
shocked fluid lead to a softening of the spectrum (less energetic electrons + lower Doppler
blueshift) and reduce the burst intensity (less energy radiated away by electrons).
The end result is a weak, soft and possibly un-detectable burst.
Therefore, efficiency considerations also require that $t^{SY} < t_{dec}$, which,
using equations (\ref{tdec}) and (\ref{cool}), leads to
\begin{equation}
\log\Gamma_{0,2}+ \frac{3}{4}\log \left( f_{\kappa}\lambda_B \right) + \frac{1}{4}\log E_{0,51}
+\frac{1}{2} \log n_0 > -1.9 \; .
\label{eff}
\end{equation}
Note that if this condition is satisfied by the electrons with the minimum random
$\gamma_m$ then it is also satisfied by the more energetic electrons (with $\gamma_e > \gamma_m$).

 The observed burst durations determine the range of $\Gamma_0$ (from eq$.$ [\ref{duration}] and
$10\,{\rm ms} \siml T_b \siml 1000\,{\rm s}$, it results that $100 \siml \Gamma_0 < 5000$),
thus equations (\ref{window}) and (\ref{eff}) can be used to constrain the energy release
parameters ($\lambda_B,\kappa$).
Numerical simulations for fireball Lorentz factors $\Gamma_0 > 1000$ require
a large computational effort, so hereafter we will restrict our attention to
cases with $\Gamma_0 < 1000$, which give burst durations $T_b \simg 1\;
{\rm s}$ (from eq$.$ [\ref{duration}]), i.e$.$ those bursts that are most often
considered in the GRB statistics. For such initial Lorentz factors,
$\kappa$ must be larger than $10^{-2}$ and $\lambda_B$ should not be less than $10^{-4}$
in order to give a spectral peak in the BATSE window.
The burst fluence in the BATSE window is determined 
also by the fraction of the available energy $E_0$ which is radiated at a power
large enough to give at detector a photon flux above a given threshold.
It would be wrong to assume
that this fraction is strictly proportional to $f_{\kappa}=\kappa/(\kappa+1)$, 
the fractional energy in electrons after shock acceleration because, even if electrons and
protons are completely ``decoupled'' after shock acceleration (i.e$.$ no further energy flow
from protons to the rapidly cooling electrons), the heat
stored in protons drives forward the FS, which accelerates new electrons. In
this indirect way a substantial fraction of proton energy can be transferred to
electrons and radiated. Numerically we found that in $\Delta t=2\,t_{dec}$ a
burst with $\kappa=0.1$ and $\lambda_B=1$ radiates $\approx 50\%$ of the total
energy $E_0$, which is not much less than the $\approx 80\%$ of $E_0$ that a burst
with $\kappa=1$ and $\lambda_B=1$ radiates during the same time. For this reason it can
considered that $\kappa$ does not have an important effect on the energy released as
long as it is not much less than $10^{-1}$.

 It is easy to see that if most of the SY-FS radiation is in the BATSE window (i.e$.$ eq$.$ [\ref{window}]
is satisfied) then either the K-N effect reduces severely the IC emission (eq$.$ [\ref{cond}]
is fulfilled) or the Kompaneets parameter $Y^{FS}$ (eq$.$ [\ref{CompFS}]) is less than 1. 
This means that if a burst observed by BATSE
represents the SY radiation emitted by the shocked EM, then the IC-FS radiation from the same
fluid is less energetic than the SY-FS emission and can be safely neglected in calculating
the cooling time-scale. On the other hand, if equation (\ref{cond}) is not satisfied 
(i.e the K-N cut-off does not reduce the efficiency of IC scattering behind the FS) 
then the SY-FS radiation does not arrive in the BATSE window. 
This suggests that a burst visible to BATSE can also be obtained 
from the IC-FS radiation if the efficiency conditions [1] $10\,{\rm keV} 
\siml E_p^{IC,FS} \siml {\rm few \; MeV}$,
[2] $Y^{FS} > 1$ and [3] $t_{cool}^{IC} < t_{dec}$
are simultaneously satisfied (condition [3] is relevant for the burst efficiency only if condition
[2] is satisfied). It can be easily shown that condition [1] implies up-scattering in 
the Thomson regime and that it cannot be fulfilled at the same time as condition [2]. In other words,
any combination of parameters $100 \siml \Gamma_0 \siml
5000,\; 10^{-2} \siml \kappa \leq 1,\; \lambda_B \leq 1,\; E_{0,51} \sim 1,\; n_0 \sim 1$ leads to
either an IC-FS component that contains a substantial fraction of the available energy but is at
energies larger than those visible to BATSE, or to an IC-FS radiation that arrives mainly in the BATSE
window but is much less energetic than the SY-FS radiation emitted by the burst, due to a small Kompaneets
parameter.
Thus the observed bursts could be IC-FS radiation only the initial fireball 
kinetic energy is much larger than $10^{50}\,{\rm ergs/sr}$,
in which case a much more energetic emission
should be detected at energies lower than the BATSE window.
Similar conclusions have been reached previously by Sari et al$.$ (1996).

In principle two other model parameter constraints can be obtained if it is
required that all electrons are confined in the shocked fluid and that the
duration $t_{acc}$ of the of the electron acceleration process is much shorter
than the corresponding SY cooling time-scale, ensuring that electrons can reach
factors $\gamma_e$ larger than the post-FS $\gamma_m$ derived above.  The former
condition requires the electron gyration radius $R_g=\gamma_e m_e c^2/eB$ to be
less than the thickness $\Delta$ ($\simg r_{dec}/\Gamma_0^2$) of the shocked
fluid shell, while the latter condition requires that $t_{acc} \sim R_g/c \ll
t^{SY}$.  It can be shown that if the inequalities (\ref{window}) and
(\ref{eff}) are satisfied then electrons are indeed confined in the shocked fluid
and are accelerated on a time-scale much shorter than the SY cooling time-scale,
so that these two conditions do not bring any new constraints on model parameters.

 The total set of model parameters used in the simulations 
below is ($E_0,\Gamma_0,n;\kappa,\lambda_B;D$), including the luminosity distance to 
source, to which the parameter $\gamma_M/\gamma_m$ could be added in some 
special cases. In the following numerical results $D=10^{28}\, {\rm cm}$; 
however the cosmological redshift effect was not accounted for specifically, 
because the spectral redshift and temporal dilation can be included in the 
model independent of the hydrodynamic simulation. Therefore, the temporal-spectral
correlations discussed below are not of cosmological origin; they are intrinsic
properties of the bursts.

\section{Results and Comparison with Observational Data}

A hardness--brightness correlation, hardness--duration anti-correlation, 
and brightness--duration anti-correlation are straightforward predictions of
these external shock models. From equations (\ref{duration}), (\ref{SYFS}) 
and the fact that the peak flux scales as $F_p \propto E_0^{2/3}\,\Gamma_0^{8/3}\, 
n^{1/3}\, D^{-2}$ we see that the fireball's initial Lorentz factor $\Gamma_0$ 
($100 \leq \Gamma_0 \siml {\rm few} \times 10^3$) is the parameter with the 
strongest influence on the spectral and temporal burst properties. If the other 
parameters have a relatively narrow range ($1 \leq E_{0,51} \siml 10$, $n_0 
\approx 1$) or are within the limiting values suggested above ($0.1 \leq \kappa 
\leq 1$, $10^{-4} \leq \lambda_B \leq 1$), then the correlations or 
anti-correlations expected among the burst parameters are due to their 
$\Gamma_0$-dependence, and are $E_p \propto F_p^{3/2}$, $E_p \propto 
T_b^{-3/2}$, and $F_p \propto T_b^{-1}$. Evidence for a
hardness--brightness correlation has been presented Mitrofanov et al$.$ (1992), 
Paciesas et al$.$ (1992), \cite{nem94}, \cite{pel94}, and \cite{mal95},
as it is implied by the hardness ratios, break energy or $E_p$ dependencies on 
the peak count rate or brightness class shown in these articles. A quantitative 
comparison is not easy as authors seldom use $F_p$ and $E_p$ in their analyses 
(or at least the same definition of the burst hardness); nevertheless it 
appears that the observed correlation is weaker than predicted above. 
The hardness--duration anti-correlation is observed by \cite{dez92} and 
Kouveliotou et al$.$ (1993) (see however \cite{band93}), while the evidence
for a brightness--duration anti-correlation is controversial (\cite{nor95}, 
Mitrofanov et al$.$ 1996); if present, it is probably far weaker than 
indicated by the above analytic scaling. Of course, a distance dispersion of
an order of magnitude, as well as a broad luminosity function (variations of
$E_0$, $\lambda_B$ and $\kappa$ parameters among bursts) and
evolutionary effects would all tend to mask such an $F_p-T_b$ 
anti-correlation through the parameter $\Gamma_0$.

Further comparison with observational data can be done using numerical results. 
Figure 1 shows spectra (computed as flux weighted averages of 10 instantaneous 
spectra, uniformly distributed within $T_b$) generated with different 
values of $\Gamma_0$ when the other parameters are held constant.  The IC 
component from the RS is shown separately while the other components can be 
distinguished in the spectrum and are identified in this figure. Note that most 
of the burst energy is in the SY component from the FS and that an important 
fraction of this energy arrives at detector in the BATSE window if the 
parameters $\lambda_B$ and $\kappa$ are close to their maximum values 
(as predicted by eq$.$ [\ref{window}]). 
The ratio $\gamma_M/\gamma_m$ is relevant for the burst 
fluence in the BATSE window only for $\Gamma_0 = 100$. The spectra cover a fairly 
broad range in energy (13-15 orders of magnitude). The burst spectral flux 
at 550 nm is $\approx 10^{-10}\,(\Gamma_0/400)^{8/3}\; {\rm ergs\;cm^{-2}\,s^{-1}
\,eV^{-1}} = 40\,(\Gamma_0/400)^{8/3}\;{\rm mJy}$, which 
corresponds to a magnitude $V \simeq 13 - 6.7\, \log\,(\Gamma_0/400)$. 

We can now estimate the effect of approximation [5] above (mixed RS--FS combinations in
the IC spectrum are neglected), using the 
previous equations for the minimum electron Lorentz factor behind each shock 
and equations (\ref{SYRS}) and (\ref{SYFS}). The energy (in the laboratory
frame) of the SY-FS photons that would be up-scattered by post-RS electrons with $\gamma_{m,RS}$ at the
limit between Thomson and K-N regimes is $E_{KN}^{RS \leftarrow FS} =
(m_e c^2/\gamma_{m,RS})\,\Gamma \simeq 1\,f_{\kappa}^{-1}\,(\Gamma_0/400)\; {\rm MeV}$.
Equation (\ref{SYFS}) and Figure 1 show that there are SY-FS photons 
less energetic than $E_{KN}^{RS \leftarrow FS}$. Therefore, due to approximation [5], a fifth component of the 
spectrum (SY-FS photons IC scattered in the RS) is neglected. This component would have a peak below 
$E_{KN}^{RS \leftarrow FS}\, \gamma_{m,RS}^2 \simeq 10\,f_{\kappa}\,(\Gamma_0/400)\; {\rm GeV}$ 
if $E_{KN}^{RS \leftarrow FS} < E_p^{SY,FS}$ or at $E_{IC}^{RS \leftarrow FS} = E_p^{SY,FS}\,\gamma_{m,RS}^2
\simeq 10\,f_{\kappa}^4\, \lambda_B^{1/2}\, n_0^{1/2}\,(\Gamma_0/400)^4\;{\rm GeV}$
if $E_{KN}^{RS \leftarrow FS} > E_p^{SY,FS}$.
The energy of the SY-RS photons that would be up-scattered by FS electrons with $\gamma_{m,FS}$ 
in a mild K-N regime is $E_{KN}^{RS \rightarrow FS} = (m_e c^2/\gamma_{m,FS})\,\Gamma \simeq 
0.8\,f_{\kappa}^{-1}\; {\rm keV}$. 
Equation (\ref{SYRS}) and Figure 1 show that there are SY-RS photons at energies lower than  
$E_{KN}^{RS \rightarrow FS}$. Approximation [5] does not take into account a sixth component of the spectrum
(SY-RS photons IC scattered in the FS) that would appear at 
$E_{IC}^{RS \rightarrow FS} = E_p^{SY,RS}\, \gamma_{m,FS}^2 = E_{IC}^{RS \leftarrow FS}$.
It can be shown that the cooling of FS electrons with $\gamma_{m,FS}$ through this kind of scatterings is
less efficient than through SY emission.
Therefore, the numerical results do not take into account the mixed components for IC scattering
and under-estimate the burst flux in the lower energy part of the IC-FS
components shown in Figure 1. Fortunately, the flux in the most important energy range (the BATSE window) 
is very little affected.
Otherwise, the intensity of the IC component relative to the SY emission from each shock, as shown
in Figure 1, is consistent with the previous estimations (eqs$.$ [\ref{CompRS}], [\ref{CompFS}],
and [\ref{CompKN}]).

The peak energy $E_p$ of the spectra shown in Figure 1 passes through the BATSE 
window as $\Gamma_0$ is increased from 100 to 800.  As expected, higher Lorentz 
factors lead to harder spectra (see legend). This can be also seen using the
hardness ratio ${\rm HR}_{32}$, defined as the ratio of counts in the third 
BATSE channel (100 keV -- 300 keV) to that in the second channel (50 keV -- 
100 keV): ${\rm HR}_{32}(\Gamma_0=100,\gamma_M/\gamma_m=100)=0.46$ ($T_b\simeq 500\,{\rm s}$),
${\rm HR}_{32}(\Gamma_0=200)=0.50$ ($T_b\simeq 100\,{\rm s}$), 
${\rm HR}_{32}(\Gamma_0=400)=0.80$ ($T_b\simeq 10\,{\rm s}$), and 
${\rm HR}_{32}(\Gamma_0=800)=0.98$ ($T_b\simeq 2\,{\rm s}$).
Figure 2 shows the SY-RS spectra obtained for a fixed $\Gamma_0=400$ and 
combinations of parameters ($n;\lambda_B,\kappa$) in which only one parameter 
is changed relative to the ``standard" combination ($1\,{\rm cm^{-3}};1,1$), 
showing the effect produced by each parameter and allowing comparison with the 
spectral peaks given by equation (\ref{SYFS}). The hardness ratios for the 
new spectra are ${\rm HR}_{32}(0.1;1,1)=0.62$, ${\rm HR}_{32}(1;0.1,1)=0.62$, 
and ${\rm HR}_{32}(1;1,0.1)=0.43$ .

The hardness ratio range allowed by the model is less wide than the range of 
$E_p$ generated by the values of $\Gamma_0$ considered, and it is useful to 
compare these ratios with those of the observed bursts. The ${\rm HR}_{32}$ 
values above are consistent to those presented by Paciesas et al$.$ (1992) 
and comparable to those found by Nemiroff et al$.$ (1994), Mitrofanov et al$.$ 
(1996), and Kouveliotou et al$.$ (1993). According to this last reference, the 
average ${\rm HR}_{32}$ is $0.87$ for bursts with $T_b > 2 \,{\rm s}$.
The hardness ratios ${\rm HR}_{43}$ of the simulated bursts range from $0.2$ to 
$ 0.4$ for $200 < \Gamma_0 < 400$ ($10\, {\rm s} < T_b < 100\, {\rm s}$) 
and is $\simeq 0.6$ for $\Gamma_0 =800$, in good agreement with the values 
calculated by Dezalay et al$.$ (1992). For the bursts shown in Figures 1 and 2
the ratio ${\rm HR}_{34,12}$ of the photon fluxes in channels 3+4 (above 100 keV) and 
in channels 1+2 (25 keV -- 100 keV) is between $0.25$ and $0.65$, lower than 
the hardness ratios calculated by Bhat et al$.$ (1994): $0.3 < {\rm HR}_{34,12} < 1$.
It is also important to compare with observations the low and high energy 
spectral indices $\alpha$ and $\beta$ as defined by Band et 
al$.$ (1993): $\alpha$ is the asymptotic limit of the slope of the photon spectrum
$C_E= dN_{\gamma}/dE$ at arbitrarily low energy ($C_E \propto E^{\alpha}\exp{[-E/E_0]}$), and
$\beta$ is the slope of $C_E$ at energies higher than the spectral 
peak $E_p$ ($C_E \propto E^{\beta}$). For the $\Gamma_0=400$ and $\Gamma_0=800$ spectra
shown in Figure 1, the spectral indices $\alpha$ obtained from fits of spectra
below $E_p$, using the Band function, are $-1.8$ and $-1.6$, respectively. 
The high energy spectral indices for the same initial Lorentz factors are
$-2.9$ and $-2.8$ . These values are consistent with those found by Band et al$.$ (1993):
$-1.5 \leq \alpha \leq 0$ and $-3 \leq \beta \leq -1$. The expected analytic value of $\alpha$
is $-1.5$ (integrated spectrum of SY radiation from cooling electrons) while that of $\beta$ is $-(1+p/2)=-2.5$
(spectrum of SY radiation from a steady-state distribution of electrons with continuous power-law injection).
The slightly lower values of the indices obtained numerically are due to the continuous
deceleration of the FS, leading to a progressive spectral softening through decreasing magnetic field, Doppler blueshift 
factor and random Lorentz factor of injected electrons. 
The increased steepness of the SY-FS spectra shown in Figure 1 below $\sim 1\;{\rm keV}$ 
is due to IC scattering in Thomson or mild KN regimes, as predicted by equation (\ref{KN}).

 A spectral evolution of GRB from hard to soft has been observed by many authors 
(e.g$.$  \cite{nor86}, \cite{mit92a}, \cite{band92}, \cite{bhat94}, 
\cite{ford95}). Figure 3 shows the light-curve and temporal evolution of the 
spectrum resulting from a simulation with constant parameters $\lambda_B$ and 
$\kappa$. 
A substantial fraction (60\%) of the burst radiation falls 
within the BATSE channels 1--4. The burst light-curve exhibits a sharp rise 
and a slow decay during which the flux is well approximated as a power-law
$F \propto T^{-1.2}$. The bottom graph shows the burst's hard to soft spectral 
evolution: the hardness ratio ${\rm HR}_{32}$, the mean energy $E_m$ in the 
BATSE channels 1--4 (defined as the ratio of the energetic flux and photon flux in 
this band), and the peak energy $E_p$ decrease monotonously during the burst 
(see legend). During the light-curve decay ($T \geq 3\; {\rm s}$), these spectral 
parameters can be approximated by power-laws in $T$: ${\rm HR}_{32} 
\propto T^{-0.1}$, $E_m \propto T^{-0.2}$ and $E_p \propto T^{-1.2}$ (similar indices
describe the spectral evolution of the other bursts shown in Figures 1 and 2). The peak
flux and spectrum of this burst show that its peak photon flux in the BATSE 
window is of order $0.1\; \gamma\,/\,{\rm cm^2 s}$, corresponding to a weak 
burst. This is due in part to the conservative choice $E_0 = 10^{51}\; {\rm 
ergs}$ over $4\pi$ steradians and to the almost maximal luminosity distance $D = 
10^{28}\; {\rm cm} \simeq 10 \; {\rm Gly}$ in this example. 
Beaming of the fireball in a 
solid angle $< 1\; {\rm sr}$ would easily boost the peak photon flux of 
this burst above $1\;\gamma\,/\,{\rm cm^2 s}$.

 In an efficient burst, the synchrotron cooling time of the FS electrons is 
much shorter than the hydrodynamic time-scale. Consequently,
most of the burst radiation is emitted by the leading edge of the expanding 
shell of shocked fluid, from a region which is $t_{dec}/t^{SY} \approx 
10^3 \div 10^4$ times thinner than the shell containing all the shocked fluid.
At detector time $T$ corresponding to $t$, the observer is not receiving radiation 
from this very thin sub-shell, but from a very elongated ellipsoid 
(see \cite{rees66}) of semi-major axis $\sim 1\,r_{dec}$. 
Consequently, the detector receives radiation that was emitted at times spread over $\sim 1\,t_{dec}$,
which means that the spectrum and light-curve reflect the long time-scale variations
of the burst physical parameters while all features
arising from short time-scale variations are 
well mixed and less distinguishable. The Lorentz factor $\Gamma(t)$
of the shocked fluid is monotonously decreasing; therefore, at constant 
energy release parameters $\kappa$ and $\lambda_B$
(relaxation of this assumption is considered in the next section), 
the spectral evolution of the burst shows only the time-evolution of $\Gamma(t)$. 
Thus, the hard to soft spectral evolution shown in Figure 3 is purely due to 
the deceleration of the radiating fluid.

\section{Burst Substructure}

We further test the ability of the blast wave model to accommodate some 
of the more frequently observed features of spectral evolution in bursts 
that exhibit individual pulses: \\
\hspace*{5mm} (i) the spectrum hardens before an 
intensity spike, and softens while the photon flux is still increasing 
(\cite{mit92b}, \cite{kou92}, \cite{band92}, \cite{bhat94}, \cite{ford95}); \\
\hspace*{5mm} (ii) the hardness of successive spikes decreases (\cite{nor86}, \cite{band92}, 
\cite{ford95}); \\
\hspace*{5mm} (iii) pulses peak earlier in the higher energy bands 
(\cite{nor86}, \cite{kou92}, \cite{nor96}); \\
\hspace*{5mm} (iv) pulses exhibit faster rises 
at higher energies and longer decays at lower energies (\cite{nor96}) and thus 
peaks are shorter at higher energy (\cite{link93}, \cite{fen95}, \cite{mit96}), \\
although exceptions from these ``rules" are not un-common. Since the flow 
Lorentz factor of the radiating shocked fluid is monotonously decreasing, the 
simple kinematics of this fluid cannot by itself produce spectra showing 
increasing hardness, nor light-curves containing peaks (assuming spherical 
symmetry), so departures from this simplest case need to be considered in
order to explain such features.

\subsection{Temporal variability from EM inhomogeneities}

The pulses that are observed in bursts could have some relation to
fluctuations in the EM density, denser EM blobs leading to a more intense
release of energy. In this scenario, the spherical symmetry is lost and a 3D
hydrodynamic code is required to perform numerical simulations.
The following is a purely analytical model of the situation.
The duration of each pulse is determined by three factors: \\
\hspace*{5mm}(1) the projection of the shocked inhomogeneity on the line of sight toward
the center of explosion, determined by: 
(1a) the laboratory frame thickness $\delta r$
of the overheated radiating region and 
(1b) the angle $\delta \, \theta =
R\,/\,r$ subtended by the shocked blob around its position $\theta$
on the spherical cap from which the observer receives radiation,
where $R$ is the radius of the un-shocked blob, assumed spherical.
$R$ should be less than the radius $r\,/\,\Gamma \sim r_{dec}\,/\,
\Gamma \simeq 3 \times 10^{13}\, n_0^{-1/3} (\Gamma_0/400)^{-5/3} \;
{\rm cm}$ of the visible spherical cap, or else the pulse lasts as long as the
whole burst, \\
\hspace*{5mm}(2) the time it takes to sweep up the entire inhomogeneity, \\
\hspace*{5mm}(3) the laboratory frame duration of the energy release $\delta t \sim t^{SY}$
(eq$.$ [\ref{cool}]).\\
The contributions of these factors to the pulse duration are: \\
\hspace*{5mm}(1a) $\Delta T_{\delta r} = \delta r/c \sim 2\,R/\Gamma_b^2 c$, \\
\hspace*{5mm}(1b) $\Delta T_{\delta \theta} =  2\,\theta R/c$,  \\
\hspace*{5mm}(2) $\Delta T_R = R\,(\Gamma_b^{-2}+\theta^2)\,/\,c$, and \\
\hspace*{5mm}(3) $\Delta T_{\delta t} = \delta t \,(\Gamma_b^{-2}+\theta^2)\,/\,2$, \\
where $\Gamma_b$ is the flow Lorentz factor of the shocked
blob, which we will approximate by the Lorentz factor $\Gamma$ of the rest of the shocked EM,
although it must be lower because the inhomogeneity is denser.
In calculating $\Delta T_{\delta r}$ above we used the fact that
in the laboratory frame the shocked blob
material is $\sim \Gamma_b^2$ times denser than before the shock,
therefore $\delta r \sim 2\,R\,/\,\Gamma_b^2$. Since $\Delta T_{\delta r}$ and
$\Delta T_R$ are of the order $R/\Gamma^2 c$ and  $\Delta
T_{\delta \theta} \approx R\,/\,\Gamma c$, it results that $\Delta T_{\delta r},\, 
\Delta T_R \ll \Delta T_{\delta \theta}$. 
Furthermore, if $R \simg 10^{10}\,\kappa^{-1}
\lambda_B^{-1} n_0^{-1} (\Gamma_0/400)^{-3}\; {\rm cm}$, then $\Delta T_{\delta t}$
can be neglected relative to $\Delta T_{\delta \theta}$.  Thus, for $10^{-3}\,
\kappa^{-1} \lambda_B^{-1} n_0^{-1} (\Gamma_0/400)^{-3} \;{\rm AU} \siml R \siml
1\, n_0^{-1/3} (\Gamma_0/400)^{-5/3} \;{\rm AU}$ (assumption 1), $\Delta T_{\delta \theta}$
determines the duration of the pulse. If $R$ is less than the lower limit set
above, then one has to consider the contribution of the cooling time to the
pulse duration. If $R$ is above the upper limit, then the pulse duration is
comparable to $T_b$ and it would be impossible to have bursts with more
than a few pulses.

 In order to derive the distribution $P(\Delta T)$ of the durations of pulses in individual
bursts, we assume that the co-moving photon number spectrum of the radiation emitted
by each blob is a power-law $dN_{\gamma} = C \varepsilon^{-\sigma} d\varepsilon$ (assumption 2)
over a range in energies $\varepsilon_{min}$ -- $\varepsilon_{max}$
wide enough that the blue-shifted corresponding laboratory frame range
covers the band in which observations are made, for all blobs that are seen by the
observer (i.e$.$ for all inhomogeneities that produce at detector a peak photon flux
above a given threshold $C_{limit}$). Thus we assume that the Doppler shifted
edges of the co-moving spectrum: $E_{min(max)}(\Gamma,\theta) = \varepsilon_
{min(max)}/[\Gamma (1-v\cos\theta)]$ ($v=\sqrt{\Gamma^2-1}/\Gamma$ is the flow velocity)
satisfy $E_{min} \leq E_m$ and
$E_M \leq E_{max}$ (assumption 3), where $E_m$ and $E_M$ are the lower and upper
edges of the observational band. If so then the peak photon flux $C_p$ of each pulse is
$C_p \propto [C/(\sigma-1)] [\Gamma (1-v\cos\theta)]^{-\sigma-2}$. The constant $C$ can
be determined using the fact that the total number of photons emitted per unit time in the
co-moving frame ($=[C/(\sigma-1)]\, \varepsilon_{min}^{1-\sigma}$, if $\sigma>1$ and $\varepsilon_{min}
\ll \varepsilon_{max}$) is equal with the number of emitting electrons $N_e$ multiplied with
the number of photons emitted per unit time by each electron, which is independent of
the Lorentz factor of the electron, and depends only on the magnetic field $B$.
If all blobs are identical not only in size but also in density (assumption 4), then $N_e$ is the same for
all pulses and therefore $C/(\sigma-1) \propto B\,\varepsilon_{min}^{\sigma-1}$.
The minimum co-moving energy $\varepsilon_{min}$ of the SY photons is proportional to $B$ and to $\gamma_m^2$,
where $\gamma_m$ is the minimum Lorentz factor of the electrons accelerated when the FS
interacts with the EM inhomogeneity. We further assume that the parameters for energy release
($\kappa,\lambda_B,\gamma_M/\gamma_m,p$) are the same for all blobs (assumption 5), so that
$B \propto \Gamma$ and $C/(\sigma-1) \propto \Gamma^{3\sigma-2}$. In the end, the peak
photon flux at detector for any pulse can be written as:
\begin{equation}
C_p(\Gamma,\theta) = \frac{const}{D^2}\, F_1(E_m,E_M;\sigma)\,
F_2 \left(p,\frac{\gamma_M}{\gamma_m};\sigma \right)\,
n_{blob}^{1+\frac{\sigma}{2}}\, \lambda_B^{\sigma/2} f_{\kappa}^{2(\sigma-1)}
R^3\, \frac{\Gamma^{2\sigma-4}}{(1-v\cos\theta)^{\sigma+2}},
\end{equation}
where $F_1$ and $F_2$ are generic notations for functions of the indicated variables and $n_{blob}$
is the density of the inhomogeneity.

 The condition $C_p(\Gamma,\theta) \geq C_{limit}$ determines which blobs yield pulses
that are detectable, provided that the kinematics $\Gamma(\Gamma_0,r)$ 
of the shell during the deceleration  phase and the spatial distribution $n_b(r)$ of the
blobs are known. For an adiabatic interaction $\Gamma(r) \propto r^{-3/2}$; numerically, we found that
$\Gamma(r)=1/2\, \Gamma_0\, (r/r_{dec})^{-3/2}$ is a good approximation. 
The number density of the EM inhomogeneities is considered to be a power-law:
$n_b(r) \propto r^{-m}$ (assumption 6), thus the homogeneous distribution is the particular case $m=0$.
Based on these assumptions one can determine for any shell position $r$ the maximum angle $\theta_{max}(r)$
relative to the line of sight toward the center of explosion for which $C_p(r,\theta_{max}[r]) =
C_{limit}$ and integrate over $r$ and $\theta$ to find the pulse duration distribution. 
Figure 4 (upper graph) shows this distribution
for a representative set of parameters ($R,\Gamma_0,m,\sigma,C_{max}/C_{limit}$), where the last parameter
is a measure of how bright is the pulse from a blob located at ($r=1\,r_{dec},\theta=0$) relative to
the detection threshold. The same figure allows one to assess the importance of each parameter:
$P(\Delta T)$ is rather insensitive to $\Gamma_0$ and depends strongly on $R$.
The lack of correlation with the initial Lorentz factor is due to the fact that $\theta_{max}(r)$
is weakly dependent on $\Gamma_0$ while the strong correlation with the blob size is clearly implied by
$\Delta T_{\delta\theta} \propto R$. If the observed burst substructure is due to EM inhomogeneities
and if the assumptions
made here are not far from reality, then the latter correlation could be used to infer from observations
the typical size of these inhomogeneities. As expected, if the pulse detection threshold is decreased, longer
pulses are seen, as more blobs at larger angles become brighter than $C_{limit}$.

 The statistics of pulses in a set of bursts can be derived by convolving the pulse duration
distribution for individual bursts with the distribution $P(\Gamma_0)$ of the initial Lorentz factors
of the shells that generated the bursts in that set. For this, we assume that $P(\Gamma_0) \propto
\Gamma_0^{\nu}$ for $\Gamma_{min} \leq \Gamma \leq \Gamma_{max}$ (assumption 7),
that all shells run into the same EM (assumption 8), and that
all bursts distances are the same (assumption 9). Such pulse duration distributions are shown in
Figure 4 (lower graph) for $\Gamma_{min}=200$ and $\Gamma_{max}=800$. $C_{max}$ (as defined above)
for $\Gamma_0=200$ was chosen 10 times larger than $C_{limit}$; this determines $C_{max}$ for any 
other $\Gamma_0$.
It can be seen that $P(\Delta T)$ is not strongly dependent on the parameters $\nu$ and $m$. Thus, it is
possible to estimate the size of the blobs by using durations of pulses in different bursts, as $R$ remains
the parameter that affects the most the pulse duration distribution.
The pulse decomposition performed by Norris et al$.$ (1996)
shows that for the brightest bursts $\Delta T$ is between $0.1\, {\rm s}$ and few seconds, therefore
$R$ must be of order $0.1 \,{\rm AU}$.

 If radiation is emitted not only from the higher density blobs but also from 
the rest of the EM, then the effect of a more intense emission of  
radiation from a blob combined with a stronger decrease of the flow Lorentz 
factor induced by the same blobs is likely to lead to a shallow peak. In other 
words, the radiating power of the source is increased but, in the same time, 
the radiation received by the detector is more stretched out in time than
the radiation emitted before and after, ironing out the peak. We are forced 
thus to assume that only the blobs emit significant radiation (perhaps
due to an enhanced magnetic field). In this case, however, a new difficulty 
arises: as \cite{sari97} pointed out, if pulses do not overlap significantly 
then the upper bound on the size of the emitting blobs set by 
the observed pulse durations limits to about 1\% the fraction of the area covered by 
these blobs on the spherical cap visible to the observer, leading to a low 
burst efficiency. 
(A higher efficiency can be reached if the number of blobs is large enough to 
cover the entire spherical cap visible by the observer, but then the pulses 
lose their individuality, producing a single hump burst.)
In order to explain the observed burst fluences, one has then
to assume that the ejecta is beamed into a fraction 1/100 of the full sky and 
that almost 99\% of the initial energy is not released as $\gamma$-rays or is 
lost adiabatically. Thus, in principle, this explanation for a complicated
pulse structure can work if the ejecta is in a jet, without increasing the 
total energy above $10^{51}$ ergs, if 99\% of this energy can go undetected.

\subsection{Temporal variability from energy release fluctuations}

To explore the limits of the ability of external shock models to generate pulses,
we consider a second, idealized scenario, in which the burst sub-structure is
due to fluctuations in the parameters $\lambda_B$ and $\kappa$ which
characterize the release of the internal energy stored in the shocked gas.
Here we consider the case where $\kappa$ is constant in time, and we assume
a variable magnetic field.
A time varying $\kappa$ should have a similar effect on the cooling time-scale
($t^{SY} \propto \lambda_B^{-1}\, \kappa^{-1}$), but a stronger one on the
spectrum ($E_p^{SY,RS} \propto \lambda_B^{1/2}\, \kappa^2$).
If the magnetic fields are such that: (1) at their maximum value the burst
radiates mainly in the BATSE window and (2) at minimum value, $t^{SY} >
t_{dec}$ (the source is in a $\gamma$-quiet phase), then equations (\ref{SYFS})
and (\ref{cool}) show that $\lambda_B$ must vary by more than 4 orders of
magnitude: $\lambda_{B,min} \siml 10^{-4}$ and $\lambda_{B,max} \siml 1$,
i.e$.$ the magnetic field must vary by at least two orders of magnitude. We do not
speculate here on the nature of the microscopic process that could produce
such fluctuations of more than 2 orders of magnitude in the magnetic field
strength over time-scales that should be shorter than $0.1\;t_{dec}$,
and remark only that plasma dynamo mechanisms which build up
the field to a fraction of the equipartition value could plausibly result in
such field variations. In the presence of such variations, multiple peaked
bursts are obtained, as shown below.

In Figure 5 we show a burst with two peaks, resulting from a relatively 
large scale variation of $\lambda_B$ (see the inset of panel a). The spectral 
evolution is shown with open symbols in panel (b): $E_m$ is decreasing
during the first peak ($T_{p,1} \simeq 1\,{\rm s}$), then increases and peaks 
around $T=2\,{\rm s}$, approximately $1\,{\rm s}$ before the photon flux and 
energetic flux peaks ($T_{p,2}\simeq 3\, {\rm s}$) and monotonously decreases 
through the remainder of the burst.  
The hardness ratio ${\rm HR}_{32}$ shows the same behavior. The monotonous 
spectral softening of the burst during the first peak is due to the deceleration
of the shocked fluid and also to the fact that this simulation
was started from $0.5\;t_{dec}$. Thus, the radiation emitted by the fluid moving at
angles $\sim \Gamma^{-1}$ (relative to the observer) prior to $t=0.5\;t_{dec}$ is not
accounted for, resulting in an artificial softening of the spectrum during the first peak 
that obscures the spectral evolution expected from a variable magnetic field. This is not
the case with the spectral evolution during second peak, which shows clearly the second
$\lambda_B$-pulse.
The duration and temporal symmetry of each peak can be characterized through 
the rise and fall times $T_{R}= \int_{0}^{T_p}dT\,f(T)$, 
$T_{F}= \int_{T_p}^{T_b}dT\,f(T)$, where $f(T)$ is the photon (or energetic) flux
normalized to its maximum value (reached at the peak time $T_p$), and through
the time-asymmetry ratio $A = T_F/T_R$.  $T_p$, the pulse duration $\Delta T 
= T_R + T_F$ and the ratio $A$ are given for each peak in the legends 
of panels (c) and (d). Note that both pulses are narrower and peak earlier at 
higher energies, which are features known to occur in observed GRBs.
The rise and fall times of the pulses decrease with energy, but their time-asymmetries show 
opposite trends: the first pulse appears more symmetric at higher energy while 
the second is more symmetric at lower energies.  In $\log \Delta T - 
\log \overline{E}$, where $\overline{E}$ is the geometrical mean of the low and 
high edges of the four BATSE channels, the two pulses appear relatively 
scattered from a straight line; nevertheless, if a power-law is fitted, then 
$\Delta T \propto E^{-0.20}$. If the pulse full width at half maximum is used,
then $\Delta T_{FWHM} \propto E^{-0.24}$. A clearer power-law dependence is 
found for the single-pulse burst shown in Figure 3: $\Delta T , \Delta 
T_{FWHM} \propto E^{-0.15}$. Norris et al$.$ (1996) decomposed 41 bright GRBs 
into pulses and found that the average full width half maximum of the pulses 
varies with energy as $E^{-0.33}$ if only the separable pulses are used, and as 
$E^{-0.38}$ for all pulses in the analyzed bursts. Therefore the pulse 
duration--energy anti-correlation of our simulated bursts is somewhat weaker 
than the observed one. The second peak in graph (a) is slightly more 
time-asymmetric than the first peak (in BATSE channels 1--4: $A_1 = 5.0$ and 
$A_2 = 5.4$); it also is wider, more shifted to later times at lower energies 
(graph c vs. graph d) and spectrally softer (as shown by ${\rm HR}_{32}$
in graph b). These are exactly the relative features observed by Norris et al$.$ (1996) 
in their pulse decomposition analysis.
The blast wave model reproduces the increase in the burst
hardness before an intensity peak but the simulated spectral hardening is weaker than
what is observed.

If radiation is emitted isotropically in the co-moving frame (as would be 
the case for a turbulent magnetic field), then the observer receives radiation 
mainly from portions of the fluid moving at angles $\siml \Gamma^{-1}$ 
relative to the line of sight.  Light emitted by such a spherical cap at time 
$t$ is spread in detector time $T$ over $\Delta T(t)= r(t)\,/\,2\, 
\Gamma^2(t)$, where $r(t)$ is the radial coordinate of the cap. Since 
the flow is ultra-relativistic, $r(t) \simeq ct$ and thus $\Delta T(t) \siml 
T_b$ (from eq$.$ [\ref{duration}]). This means that any instantaneous 
event that occurs in the spherical shell is seen by the observer smeared over 
a good fraction of the burst. Pulse-like emission of radiation and spectral 
features due to a change in the fluid physical parameters are ironed out very 
efficiently by sampling over the entire opening angle of the region seen by the 
observer. This naturally suggests that, if spherical symmetry in the laboratory
frame is maintained, then the angular opening of the cap from which the detector
receives radiation must be less than $\Gamma^{-1}$ in order to reduce the blending
of the temporal and spectral features arising from fluctuations in the burst physical
parameters. This could happen if 
the radiation, instead of being emitted isotropically in the co-moving frame, 
is beamed along the radial direction of fluid motion. 
If this radiation is concentrated in two cones of solid angles $2\pi 
(1-\mu_{co})\; {\rm sr}$ around the radius vector, then the observer receives 
radiation from a cap of angular opening $[(1-\mu_{co})/(1+\mu_{co})]^{1/2}\, 
\Gamma^{-1} < \Gamma^{-1}$. 

The effect of such an anisotropic emission can be assessed from Figure 6: as 
the radiation in the co-moving frame is emitted within a narrower solid angle, 
the light-curve becomes more time-symmetric. Due to the monotonous spectral 
softening ($\lambda_B$ and $\kappa$ are constant, $\Gamma$ decreases), the 
photon flux decays more slowly than the energetic flux and therefore is more 
time-asymmetric (see the rise and fall times given in the legend of each graph).
This figure can be compared with similar ones presented by Mitrofanov et al$.$
(1996), showing the GRB ``average curve of emissivity'' in the BATSE channels 
2+3. In the isotropic case, the radiation emitted by the fluid moving at large 
angles ($\siml \Gamma^{-1}$) relative to the line of sight is Doppler 
blue-shifted by a factor $\siml 2$ relative to the radiation emitted by the fluid 
moving exactly toward the observer. This large angle radiation arrives later at 
the detector and is mixed with the radiation emitted at later times, but 
from regions moving at smaller angles. As the co-moving frame solid angle in 
which radiation is emitted decreases, the detector receives less radiation from 
the fluid moving at large angles, therefore the radiation emitted at different 
times is less mixed and the spectrum reflects better the instantaneous physical 
conditions of the radiating fluid. For Figure 6 this means that the spectrum 
shows better the deceleration of the shocked fluid in the anisotropic case than 
in the isotropic one. This can be seen in the
evolution of the three spectral parameters used so far during the burst fall ($T>T_p$): 
${\rm HR}_{32} \propto T^{-0.1}$, $E_m \propto T^{-0.2}$
and $E_p \propto T^{-1.2}$ in the isotropic case, 
${\rm HR}_{32} \propto T^{-0.5}$, $E_m \propto T^{-0.7}$ and $E_p \propto T^{-2.1}$
if in the co-moving frame the radiation is emitted 
within $4\pi/5\;{\rm sr}$ around the radial direction
while in the most anisotropic emission considered here
($4\pi/17\;{\rm sr}$ around the direction of flow) 
${\rm HR}_{32} \propto T^{-0.9}$, $E_m \propto T^{-1.3}$
and $E_p \propto T^{-2.7}$. During the burst fall, the Lorentz factor of the leading edge
of the expanding gas (from where comes most of the radiation received by the detector if 
$t^{SY} \ll t_{dec}$) is approximately $\Gamma \propto T^{-2/3}$, which implies
that the fastest possible spectral peak evolution is $E_p (\propto \Gamma^4) \propto T^{-2.7}$.
Therefore, the most anisotropic case considered above yields a spectral evolution that reflects very well
the deceleration of the shocked fluid.
For Figure 5, the anisotropic emission
allows the spectrum to show better the effect of a varying $\lambda_B$ (graph b, 
filled symbols): note the much faster spectral softening during the first pulse 
and the sharp spectral hardening before the second peak.

The same type of anisotropic emission can be used to generate multi-peak 
bursts, as shown in Figure 7. The standard of comparison is an isotropic 
emission case (top graph of Figure 7), for which the radiation coming from the 
source is blended into a single hump light-curve. The more anisotropic the
emission is, the shorter and brighter the pulses are and individual 
peaks can be distinguished better.  The progressive spectral softening makes 
these peaks to be less well separated in photon flux than in energy flux, as 
can be seen in the middle graph. Pulses appear more distinct
in the case of maximum anisotropy considered here (bottom 
graph). If much internal energy were to accumulate in the shocked fluid 
between two consecutive $\lambda_B$-pulses and if most of it is radiated 
during a magnetic field pulse, then the observed peaks may be blended into a single one
(as it happens with the pairs of pulses 1-2, 5-6 and 7-8 in the middle graph). 
The pulse onset times, calculated from the time when $t^{SY} < 
t_{dec}$ and using the radial coordinate of the shell's leading edge, are 
indicated with numbers. The peak of each pulse occurs slightly later due 
to the angular opening and thickness of the source. Note that later 
pulses are more 
time-asymmetric than earlier ones and last longer; 
this is caused by the continuous deceleration of 
the source. A larger number of pulses can be simulated by using an even 
stronger co-moving frame anisotropy.

\section{Conclusion}

We have discussed some features of numerically simulated GRB spectra and 
light-curves from external shock models, with particular attention to
the expected spectral-temporal correlations and the expected degree of
temporal substructure. The values of the most important model parameters 
($\Gamma_0,n;\lambda_B,\kappa$) were chosen such that the burst releases 
an important fraction of its energy in the BATSE window. No effort was 
made to optimize these parameters so that the simulated bursts mimic the 
observed ones, other than considering (phenomenologically) the effects of a
variable magnetic field and an anisotropic emission pattern in the co-moving 
frame for some of the models.
We then compared the features of the numerical bursts 
with those characteristics of the observed GRBs that are well established, such
as the spectral hardnesses, low and high energy spectral indices, hard-to-soft
spectral evolution, correlation between spectral hardness and intensity, 
pulse duration vs. energy etc.  We summarize here the features of the 
numerically simulated model bursts :
\begin{enumerate}
\vspace*{-4mm}
\item The brightness and spectral hardness are correlated.
\vspace*{-2mm}
\item They show a spectral hardness -- duration anti-correlation :
$E_p \propto T_b^{-3/2}$. The observed dependence is weaker, which could 
be due to the variations of the energy release parameters ($\lambda_B,\kappa$) 
from one burst to another,
\vspace*{-2mm}
\item For single pulse light-curves, the photon flux in the BATSE window 
rises as $T^{1.6}$ and decays approximately as $T^{-1.0}$.  
The fall is steeper when the co-moving frame emission is anisotropic,
\vspace*{-2mm}
\item The low energy index of the averaged spectrum is $\alpha 
= -1.7 \pm 0.2$, not far from the expected value of $-1.5$ . For $p=3$, the
high energy index is $\beta=-2.8 \pm 0.1$, not far from $-2.5$, the theoretical value. 
The former index is determined by the evolution of the 
accelerated electrons if the spectral peak $E_p$ is in the BATSE window, while the
latter index depends on the choice of the electron power-law index $p$, 
\vspace*{-2mm}
\item The spectra show a general hard to soft evolution outside of intensity 
pulses. In single hump light curves arising from isotropic co-moving frame 
emission, the spectral evolution at $T > T_p$ is characterized by
$E_p \propto T^{-1.1 \pm 0.1}$. If in the co-moving frame radiation is 
emitted preferentially on the radial direction of motion, the spectral evolution
is faster,
\vspace*{-2mm}
\item The peak (or break) energy $E_p$ increases with intensity 
during a pulse, but peaks earlier. The mean energy $E_m$ in the BATSE window and the hardness 
ratio ${\rm HR}_{32}$ (or similar ones) show a similar trend. The increase in the burst hardness
before an intensity peak is stronger in the anisotropic emission case,
\vspace*{-2mm}
\item Earlier pulses are harder and have a more time-symmetric profile at 
higher energies. Later pulses may show an opposite trend: more symmetry at 
lower energies,
\vspace*{-2mm}
\item Pulses peak earlier and are shorter in higher energy bands than at 
lower energies.  Numerically, we found that the pulse duration scales as 
$E^{-0.20 \pm 0.05}$, which is a weaker dependence than observed 
($E^{0.3 \div 0.4}$).
We must recognize here that, taking into account some of the approximations 
made, the BATSE channels are relatively narrow for the accuracy of our 
simulations, so that the calculated pulse duration vs. energy dependence 
can be considered satisfactorily close to what is observed,
\vspace*{-2mm}
\item The angular opening of the region from which the observer receives 
radiation limits the number of separate pulses to very few. A larger number 
of pulses results if radiation is not emitted isotropically in the co-moving 
frame. Later pulses are more time-asymmetric than earlier ones and last 
longer if they result from a periodic variation of the source radiating power.
\vspace*{-3mm}
\end{enumerate}

The above list of burst characteristics is in agreement with, or at least close 
to, what is observed.  It is worth noting that the brightness -- duration 
anti-correlation induced by the fireball Lorentz factor $\Gamma_0$ will be
weakened by any dispersion in some of
other parameters involved in the model, such as the distance $D$ to the burst, 
the source initial kinetic energy $E_0$, and the energy release parameters
($\kappa,\lambda_B$), which could explain why this 
anti-correlation is controversial or, at best, a very weak one. 

Another major observational feature against which to contrast models 
is the bimodality in duration distribution. One reason why this is expected
in external shock models of GRB (\cite{sari96}) is related to the limited 
energy range in which BATSE is sensitive: significant energy arrives at the
detector in the BATSE window from either the forward shock (FS), in the case 
of the long bursts, or from the reverse shock (RS), in the case of the short 
bursts, but from neither of these shocks for bursts with durations $T \simeq 
2\; {\rm s}$. Moreover, such different burst origins can explain the lack of a 
duration -- brightness anti-correlation: the RS is less efficient than the FS 
in converting the fireball kinetic energy into gamma-rays, diminishing the 
brightness of the short burst. 
In our model, the RS is always mildly relativistic and radiates inefficiently 
(i.e$.$ at energies outside the BATSE window). If the expanding shell thickness
increases faster than we considered, before its deceleration becomes important,
then the density of the colliding shell can be small enough to lead to the 
formation of a more relativistic RS. The electron Lorentz factor can be further boosted
if an injection fraction well below unity is assumed. In this case the bursts duration bimodality would be 
reproduced numerically. A different explanation (\cite{mesz93a}) 
for a bimodal duration distribution may be that shorter GRB arise from events
in a relatively dense external environment (the external shocks occur in the 
progenitor's own pre-ejected wind or in a denser disk galactic disk 
environment) while longer GRB could be due to events in a much lower density 
environment (e.g$.$ the object has moved out of its own pre-ejected wind or 
it has escaped the galactic disk).

In summary, the external shock or blast wave model can explain the spectral 
features and correlations of most bursts. It can also explain the time 
histories of those bursts which have a simple structure (up to 4-8
pulses) if the magnetic field is variable and the co-moving emissivity
is appreciably anisotropic. It is difficult to see how this could
be extended to fit also bursts with more than 8-10 pulses. There is no
difficulty in explaining the latter in outflows with ``internal" shocks (e.g$.$
\cite{rees94}), which are expected to have similar spectral properties without 
limitations on the degree of variability. Nevertheless, external shock models
show a remarkable degree of qualitative agreement with a large range of medium 
to long time-scale spectral and temporal correlations exhibited by the GRB 
data. This suggests either that external shocks may be responsible
for part of the emission of a GRB, or else that a substantial subset of bursts
(i.e$.$ the less variable ones) may be ascribed to external shock events.

\acknowledgements{This research has been supported in part through NASA NAG 5-2362.}


\input psfig

\begin{figure}
\centerline{\psfig{figure=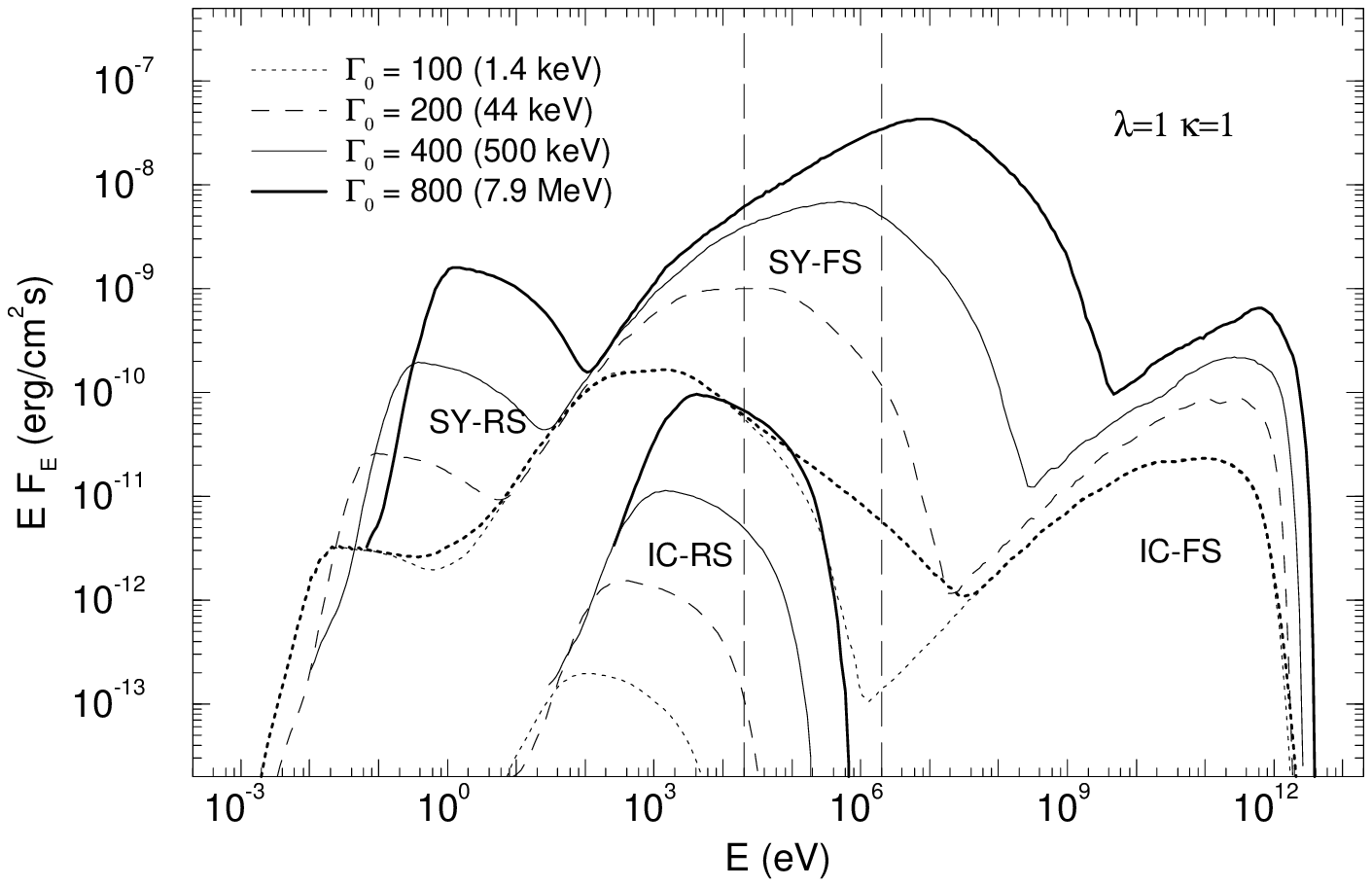}}
\vspace{1 in}
\figcaption{SY and IC spectra for $E_0=10^{51}\,{\rm ergs}$, $n=1\,{\rm cm}^{-3}$, $D=10^{28} \;{\rm cm}$,
$\lambda_B=1$, $\kappa=1$, $\gamma_M/\gamma_m=10$, $p=3$, and different parameters $\Gamma_0$.
The thick dotted curve is for $\Gamma_0=100$ and $\gamma_M/\gamma_m=100$. Labels
indicate the origin of each component: SY = synchrotron, IC = inverse Compton scattering,
RS = reverse shock, FS = forward shock. $E\,F_E=\nu\,F_{\nu}$ is the
power per logarithmic energy (or frequency) interval. Vertical dashed lines show the BATSE window.
The legend also gives the spectral peak energy $E_p$ for each parameter $\Gamma_0$.}
\end{figure}

\begin{figure}
\centerline{\psfig{figure=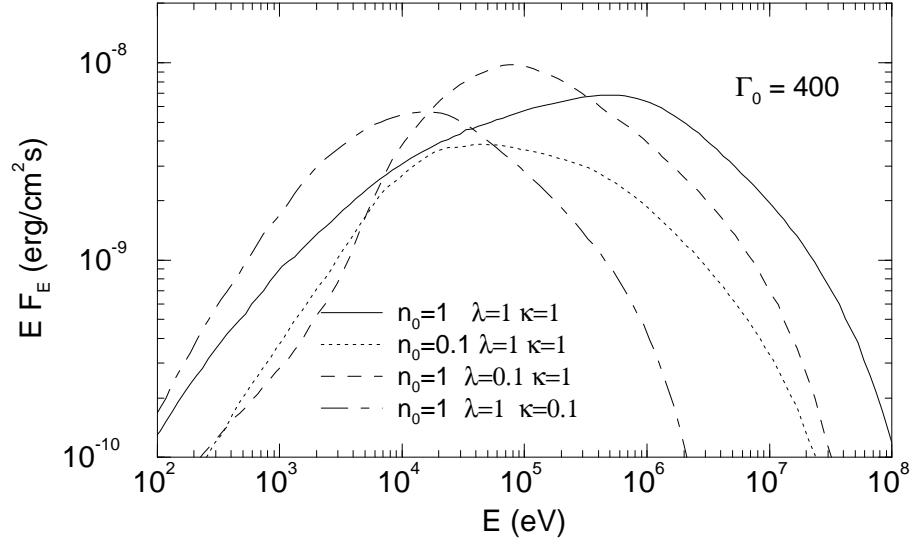}}
\vspace{3 in} 
\figcaption{SY spectra from the FS for different sets of parameters ($n_0;\lambda_B,\kappa$).
$E_0=10^{51}\,{\rm ergs}$, $\Gamma_0=400$, $D=10^{28} \;{\rm cm}$, $\gamma_M/\gamma_m=10$, and $p=3$. 
$E_p(1;1,1)=500\,{\rm keV}$,
$E_p(0.1;1,1)=45\,{\rm keV}$, $E_p(1;0.1,1)=75\,{\rm keV}$, $E_p(1;1,0.1)=17\,{\rm keV}$.} 
\end{figure}

\begin{figure}
\centerline{\psfig{figure=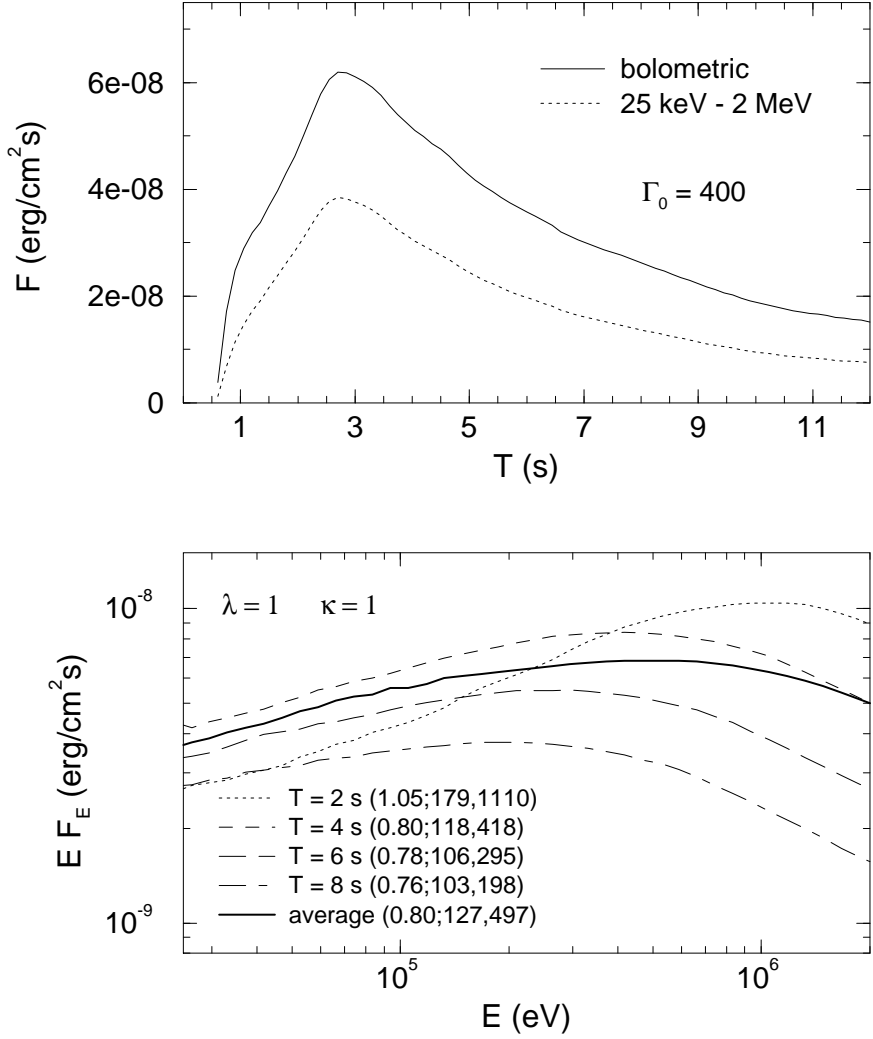}}
\vspace{1 in}
\figcaption{Light-curve (upper graph) and spectral evolution (lower graph) for $E_0=10^{51}\,{\rm ergs}$,
$n=1\,{\rm cm}^{-3}$, $\Gamma_0=400$, $D=10^{28}\;{\rm cm}$, $\lambda_B=1$, and $\kappa=1$. The detector
time $T$ is measured from the moment of the explosion that generated the fireball.
60\% of the burst energy arrives at detector in the BATSE channels 1--4. The legend of the bottom 
graph indicates the hardness ratio ${\rm HR}_{32}$, the mean energy $E_m$ in keV in the BATSE window and
the spectral peak $E_p$ in keV (in this order), at different moments and for the averaged spectrum.}
\end{figure}

\begin{figure}
\centerline{\psfig{figure=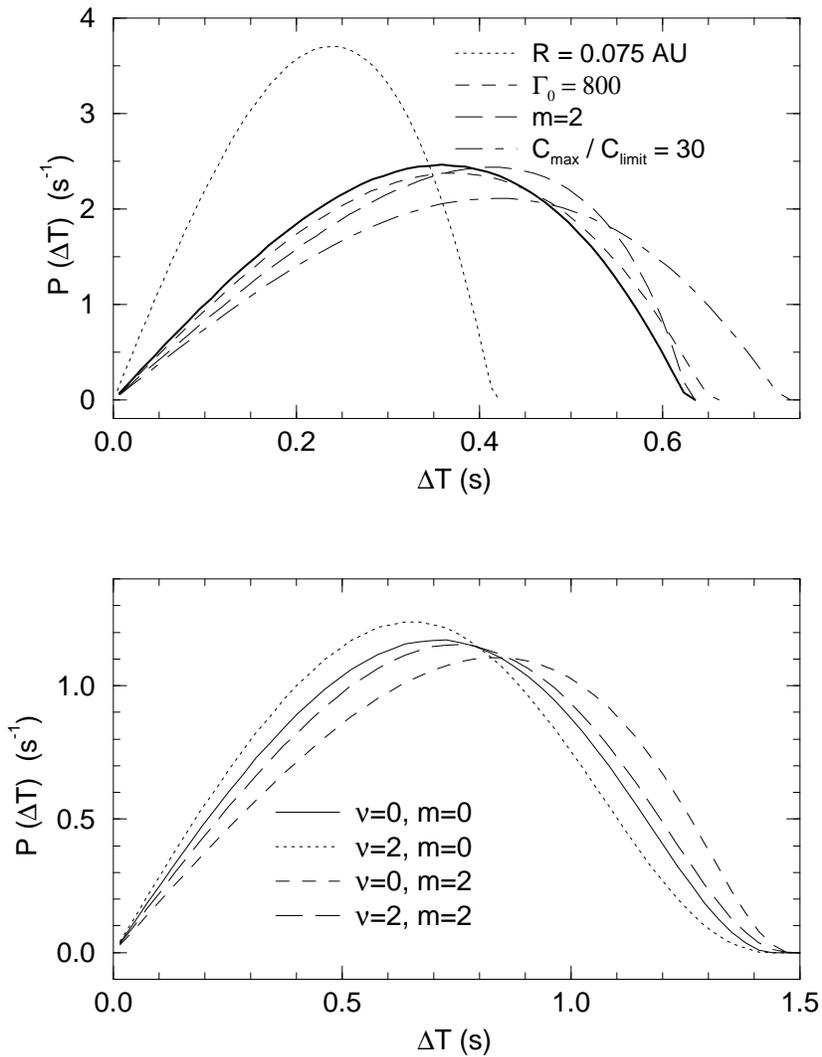}}
\vspace{1 in}
\figcaption{Burst sub-structure from EM inhomogeneities. 
Top graph: pulse duration distribution for individual bursts.
The solid thick curve is for $R=0.1\,{\rm AU}$, $\Gamma_0=400$, $\sigma=2$,
$m=0$ (homogeneous distribution of blobs), and a pulse detection threshold $C_{limit}=C_{max}/10$,
where $C_{max}$ is the photon flux that a blob located on the line of sight toward the
center of explosion and at $r=1\,r_{dec}$ yields at detector. Other distributions shown are for the
same set of parameters except that indicated in the legend.
Bottom graph: duration distribution for a set of bursts. The combinations of initial Lorentz factor
distribution and spatial distribution of the EM inhomogeneities are indicated in the legend.
Parameters: $R=0.1\,{\rm AU}$, $\Gamma_{min}=200$, $\Gamma_{max}=800$, $\sigma=2$, 
$C_{max}(\Gamma_{min})/C_{limit}=10$.}
\end{figure}

\begin{figure}
\centerline{\psfig{figure=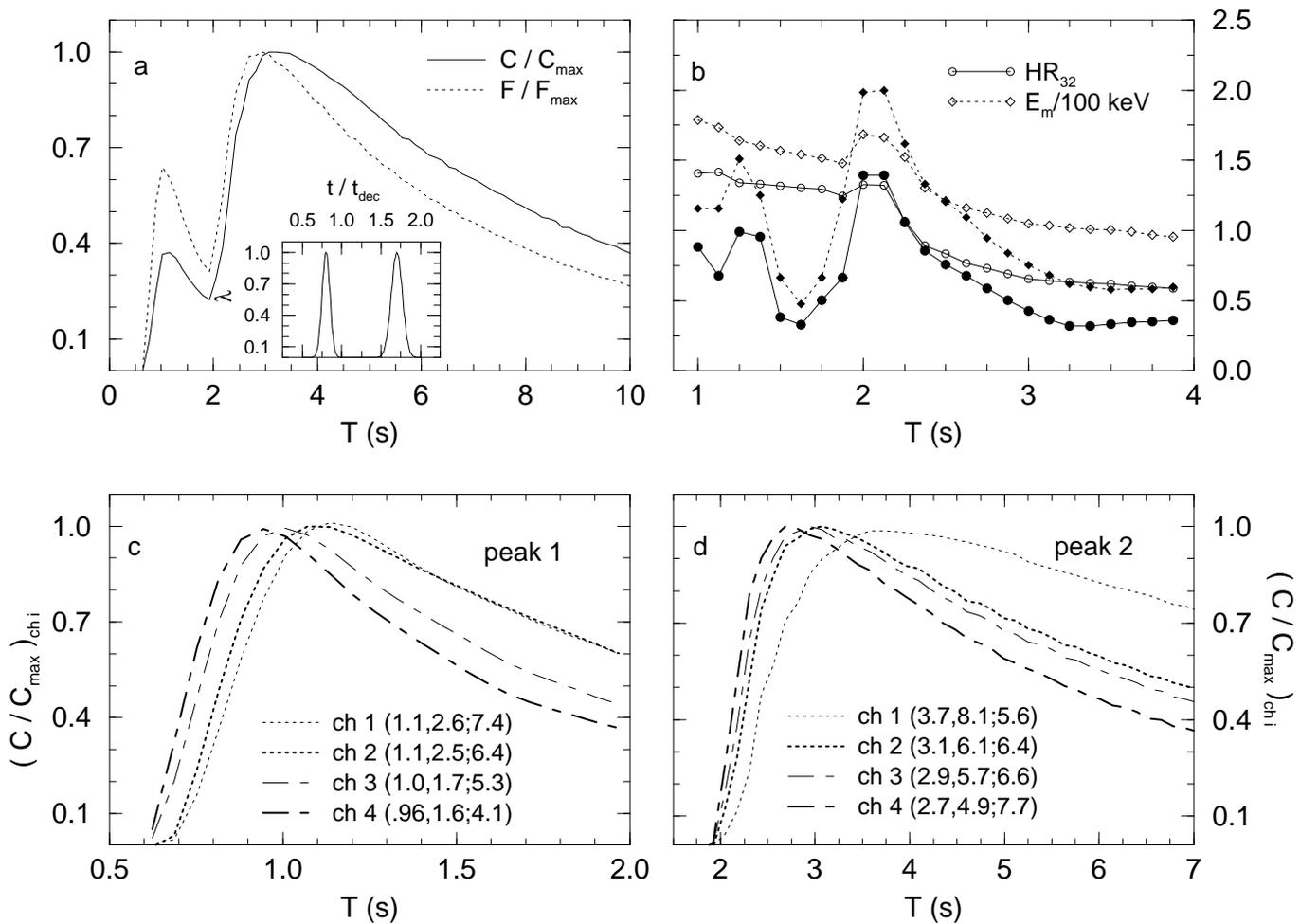}}
\vspace{0.5 in}
\figcaption{Burst sub-structure from energy release fluctuations.
Time history, spectral evolution, and pulses shapes in BATSE channels 1--4 for $E_0=10^{51}
\,{\rm ergs}$, $n=1\,{\rm cm}^{-3}$, $\Gamma_0=400$, $D=10^{28}\;{\rm cm}$, $\kappa=1$, and time variable
$\lambda_B$, shown in the inset of graph (a) (light-curve). 
(b) Hardness ratio ${\rm HR}_{32}$ and mean energy $E_m$
in the BATSE channels 1--4. Open symbols are for an isotropic emission in the co-moving frame, filled 
symbols are for an anisotropic case: radiation emitted within $4\pi/17$ sr of the radial flow direction. 
Note that the second peak ($T_p \simeq 3\,{\rm s}$) shows a stronger increase in spectral hardness  
in the latter case and that in both cases the maximum spectral hardness occurs $\sim 0.5-1\,{\rm s}$
before the intensity peak. (c) The first peak as seen in each
BATSE channel. (d) The second peak in the same bands. Fluxes in (c) and (d) are normalized to the 
peak value in that channel. Legends in (c) and (d) give the peak time, the duration (the sum of the
rise and fall times, as defined in text), and the
time-asymmetry ratio of each peak, in this order, in each BATSE channel.}
\end{figure}

\begin{figure}
\centerline{\psfig{figure=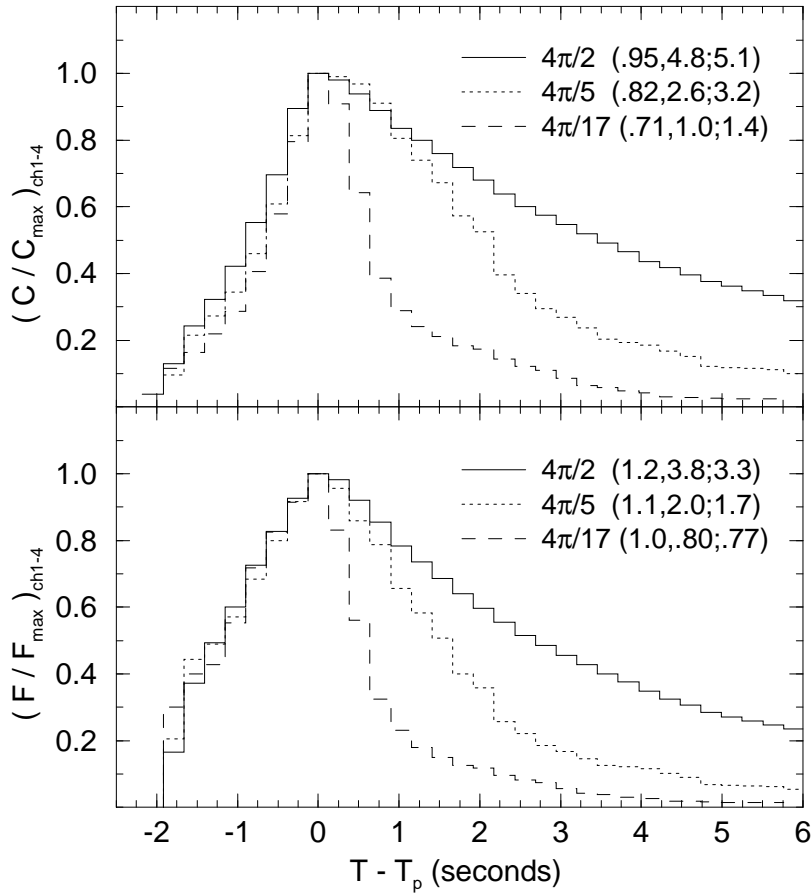}}
\vspace{1 in}
\figcaption{BATSE window time histories of a burst with $\Gamma_0=400$ and different
degrees of anisotropy of radiation emission in the co-moving frame. Solid curves: isotropic emission (two cones,
each of $4\pi/2\; {\rm sr}$ solid angle, corresponding to a cap of angular opening $\Gamma^{-1}$
in the laboratory frame). Dotted and dashed curves: anisotropic emission ($4\pi/5\; {\rm sr}$ and $4\pi/17\; 
{\rm sr}$ co-moving frame solid angles, corresponding to $0.5\, \Gamma^{-1}$ and $0.25\,
\Gamma^{-1}$ angular opening caps, respectively). Note that, as the co-moving frame anisotropy 
increases, the light-curves becomes more symmetric. Legends indicate the rise and fall times and the
asymmetry ratio, as defined in text, in this order. Upper graph: photon fluxes, lower graph: energetic fluxes.
Light-curves have been normalized to their peak fluxes, binned in 256 ms, and aligned at their peaks.}
\end{figure}

\begin{figure}
\centerline{\psfig{figure=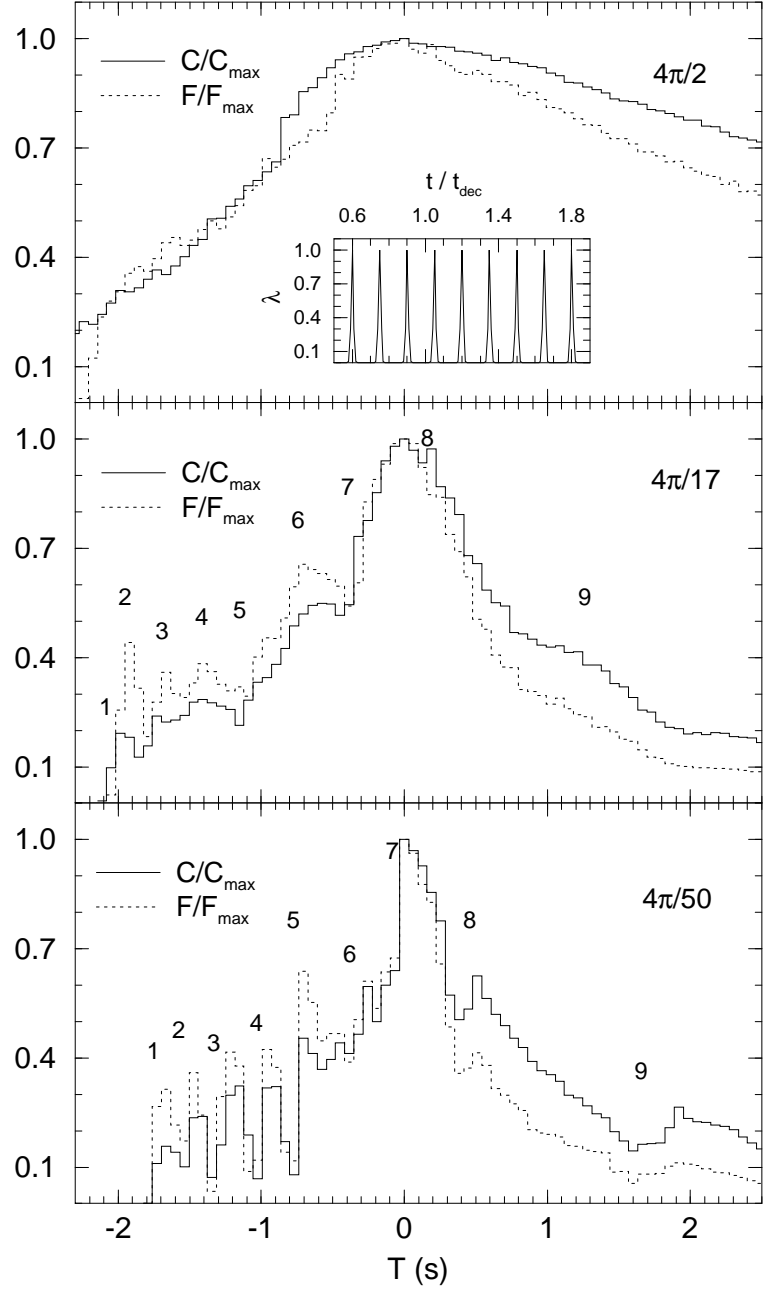}}
\vspace{1 cm}
\figcaption{$\Gamma_0=400$ multi-peak burst (in BATSE window) generated by a variable magnetic field 
(inset of upper graph).  Upper graph:
isotropic emission. Middle and lower graph: anisotropic emission. The solid angle around the direction 
of motion in which the emission
is confined is indicated in each graph. Note that as the degree of anisotropy
increases, the peaks become more distinct. Photon and energy fluxes have been normalized at
the maximum value reached during the burst, aligned at
their peak times, and binned in 64 ms. Numbers indicate the expected beginning of each pulse.}
\end{figure}

\end{document}